\documentclass[reqno]{amsart}

\usepackage{amsmath,amsfonts}
\usepackage{epsfig}
\usepackage{graphicx}
\usepackage[active]{srcltx}

\def\HN{H_N} 
\def\tpsi{\widetilde{\psi}}
\def\VN{V_N} 

\def\e{\epsilon}

\def\Tor{\mathbb T}
\def\tr{{\rm Tr}}

\def\bra{\big\langle}
\def\ket{\big\rangle}
\def\Bra{\Big\langle}
\def\Ket{\Big\rangle}

\def\N{{\mathbb N}}

\def\R{{\mathbb R}}

\def\up{{\underline{p}}}

\def\uq{{\underline{q}}}

\def\ut{{\underline{t}}}
\def\uu{{\underline{u}}}

\def\ux{{\underline{x}}}

\def\cH{{\mathcal H}}
\def\cK{{\mathcal K}}
\def\cL{{\mathcal L}}
\def\cU{{\mathcal U}}

\def\1{{\bf 1}}

\def\eqnn{\begin{eqnarray*}}
\def\eeqnn{\end{eqnarray*}}
\def\eqn{\begin{eqnarray}}
\def\eeqn{\end{eqnarray}}

\def\prf{\begin{proof}}
\def\endprf{\end{proof}}

\theoremstyle{plain}
\newtheorem{theorem}{Theorem}[section]

\newtheorem{proposition}[theorem]{Proposition}

\newtheorem{lemma}[theorem]{Lemma}
\newtheorem{corollary}[theorem]{Corollary}

\numberwithin{equation}{section}

\begin{document}

\parskip=8pt

\title[Derivation of quintic NLS]
{The quintic NLS as the mean field limit of a Boson gas 
with three-body interactions}
\author[T. Chen]{Thomas Chen}
\address{T. Chen,  
Department of Mathematics, University of Texas at Austin.}
\email{tc@math.utexas.edu}

\author[N. Pavlovi\'{c}]{Nata\v{s}a Pavlovi\'{c}}
\address{N. Pavlovi\'{c},  
Department of Mathematics, University of Texas at Austin.}
\email{natasa@math.utexas.edu}


\begin{abstract}
We investigate the dynamics of a boson gas with three-body interactions
in dimensions $d=1,2$. We
prove that in the limit of infinite particle number,  
the BBGKY hierarchy of $k$-particle marginals converges to a limiting
(Gross-Pitaevskii (GP))
hierarchy for which we prove existence and uniqueness of solutions.
Factorized solutions of the GP hierarchy are shown to be 
determined by solutions of a quintic
nonlinear Schr\"odinger equation.
Our proof is based on, and extends, methods of Erd\"os-Schlein-Yau,
Klainerman-Machedon, and Kirkpatrick-Schlein-Staffilani.
\end{abstract}

\maketitle

\section{Introduction}

In this paper, we study the dynamical mean field limit of a nonrelativistic  
Bose gas with 3-particle interactions in space dimensions $d=1,2$. 
We prove that the BBGKY hierarchy of marginal 
density matrices converges to an infinite
hierarchy whose solutions are determined by solutions of a {\em quintic} 
nonlinear Schr\"odinger equation (NLS),
provided that the initial conditions
have product form. 
In our proof, we adapt methods of Erd\"os-Schlein-Yau, \cite{esy},
Klainerman-Machedon, \cite{klma}, and
Kirkpatrick-Schlein-Staffilani, \cite{kiscst}, to this problem. Some
parts of our exposition follow quite closely \cite{kiscst} and \cite{klma}.
In a companion paper, we discuss the Cauchy problem for the GP hierarchy
in more generality, \cite{chpa2}.
 
We consider a system of $N$ bosons with Hamiltonian
\eqn\label{eq-def-Ham-0}
	\HN \, := \, \sum_{j=1}^N (-\Delta_{x_j}) \, + \, \frac{1}{N^2}
	\sum_{1\leq i < j < k \leq N} N^{2d\beta} V(N^\beta(x_i-x_j),N^\beta(x_i-x_k)) \,,
\eeqn
on the Hilbert space $\cH_N=L^2_{sym}(\R^{dN})$. The functions
$\Psi(x_1,\dots,x_N)\in\cH_N$ are fully symmetric with respect to permutations of the
arguments $x_j$.
We assume that the translation-invariant three-body potential $V$ has the properties
\eqn
	V \, \geq \, 0
	\; \; \; \; , \; \; \; \;
	V(x,y) \, = \, V(y,x)
	\; \; \; \; , \; \; \; \; 
	V \, \in \, W^{2,p}(\R^{2d})
\eeqn 
for $2d<p\leq\infty$. 
We note that, since evidently,
\eqn
	U(x_1-x_2,x_2-x_3,x_1-x_3)
	& = & U(x_1-x_2,-(x_1-x_2)+(x_1-x_3),x_1-x_3)
	\nonumber\\
	& \equiv & V(x_1-x_2,x_1-x_3) \,,
\eeqn
every translation invariant three-body 
interaction potential $U$ can be written in the above form.

The solutions of the Schr\"odinger equation
\eqn
	i\partial_t \Psi_{N,t} \, = \, \HN \, \Psi_{N,t}
\eeqn
with initial condition $\Psi_N\in\cH_N$
determine the $N$-particle density matrix
\eqn
	\gamma_{N}(t;\ux_N;\ux_N') \, = \, \overline{ \Psi_{N,t} (\ux_N) } \Psi_{N,t}(\ux_N')
\eeqn 
and its $k$-particle marginals
\eqn
	\gamma_{N,t}^{(k)}(t;\ux_k;\ux_k') \, = \, 
	\int d\ux_{N-k} \gamma_{N}(t;\ux_k,\ux_{N-k};\ux_k',\ux_{N-k}) \,,
\eeqn
for $k=1,\dots,N$, where $\ux_k=(x_1,\dots,x_k)$, $\ux_{N-k}=(x_{k+1},\dots,x_N)$, etc.

The BBGKY hierarchy is given by
\eqn
	\lefteqn{
	i\partial_t \gamma_{N,t}^{(k)} \, = \, \sum_{j=1}^k [-\Delta_{x_j},\gamma_{N,t}^{(k)}]
	+ \frac{1}{N^2}\sum_{1\leq i<j<\ell\leq k} [\VN(x_i-x_j , x_i-x_\ell) , \gamma_{N,t}^{(k)} ] 
	}
	\nonumber\\
	&&+\frac{(N-k)}{N^2}
	\sum_{1\leq i<j\leq k}\tr_{k+1} [\VN(x_i-x_{j}, x_i-x_{k+1}) , \gamma_{N,t}^{(k+1)} ]
	\\ \label{intro-BBGKY}
	&&+\frac{(N-k)(N-k-1)}{N^2}
	\sum_{j=1}^k\tr_{k+1}\tr_{k+2}[ \VN( x_j-x_{k+1}, x_j-x_{k+2} ) , \gamma_{N,t}^{(k+2)} ]
	\nonumber
\eeqn
where
\eqn
	\VN(x,y) \, := \, N^{2d\beta} V(N^\beta x, N^\beta y) \,.
\eeqn
We note that in the limit $N\rightarrow\infty$, the sums weighted
by combinatorial factors have the following size. In
the first interaction term on the
rhs, we have $\frac{k^2}{N^2}\rightarrow0$ for every fixed $k$
and sufficiently small $\beta$,
and for the second term $\frac{(N-k)k}{N^2}\approx\frac{k}{N}\rightarrow0$.
For the third interaction term on the rhs, we note that $\frac{(N-k)(N-k-1)}{N^2}\rightarrow1$
for every fixed $k$.
Accordingly, a rigorous argument outlined in Section \ref{sect-compactness-1}
shows that in the limit $N\rightarrow\infty$, one obtains the infinite
hierarchy
\eqn\label{eq-def-b0-1}
	i\partial_t \gamma_{\infty,t}^{(k)} & = & \sum_{j=1}^k [-\Delta_{x_j},\gamma_{\infty,t}^{(k)}]  
	\, + \, b_0 \, \sum_{j=1}^kB_{j;k+1,k+2} \gamma_{\infty,t}^{(k+2)}
\eeqn
where 
\eqn
	b_0 \, = \, \int dx_1 \, dx_2 \, V(x_1,x_2)
\eeqn
is the coupling constant,
and where we will sometimes refer to
\eqn
	B_{j;k+1,k+2}\gamma_{\infty,t}^{(k+2)}
       := \, B^{+}_{j;k+1,k+2}\gamma_{\infty,t}^{(k+2)} - 
                    B^{-}_{j;k+1,k+2}\gamma_{\infty,t}^{(k+2)}, 
\nonumber
\eeqn 
where
\eqn
	\lefteqn{
	\left( B^{+}_{j;k+1,k+2}\gamma_{\infty,t}^{(k+2)} \right) (x_1,\dots,x_k;x_1',\dots,x_k')
	}
	\nonumber\\
	&& := \, \int dx_{k+1}dx_{k+1}'dx_{k+2}dx_{k+2}' \,
	\nonumber\\
	&&\quad\quad\quad\quad 
	\delta(x_j-x_{k+1})\delta(x_j-x_{k+1}')\delta(x_j-x_{k+2})\delta(x_j-x_{k+2}')
	\nonumber\\
       && \quad\quad\quad\quad\quad\quad 
	\gamma_{\infty,t}^{(k+2)}(x_1,\dots,x_{k+2};x_1',\dots,x_{k+2}')
	\nonumber\, 
\eeqn

and 

\eqn
	\lefteqn{
	\left( B^{-}_{j;k+1,k+2}\gamma_{\infty,t}^{(k+2)} \right)(x_1,\dots,x_k;x_1',\dots,x_k') 
	}
	\nonumber\\
	&& := \, \int dx_{k+1}dx_{k+1}'dx_{k+2}dx_{k+2}' \,
	\nonumber\\
	&&\quad\quad\quad\quad 
	\delta(x_j'-x_{k+1})\delta(x_j'-x_{k+1}')\delta(x_j'-x_{k+2})\delta(x_j'-x_{k+2}')
	\nonumber\\
	&& \quad\quad\quad\quad\quad\quad 
	\gamma_{\infty,t}^{(k+2)}(x_1,\dots,x_{k+2};x_1',\dots,x_{k+2}')
	\nonumber\, 
\eeqn
as the ``contraction operator".
The topology in which this convergence holds is described in
Section \ref{sect-compactness-1} below, and is here adopted from \cite{esy,kiscst}.

Written in integral form, 
\eqn\label{eq-BBGKY-intform-1} 
	\gamma_{\infty,t}^{(k)} \, = \, \cU^{(k)}(t) \, \gamma_{\infty,0}^{(k)} 
	- \, i \, b_0 \, \sum_{j=1}^k \int_0^t ds \, \cU^{(k)}(t-s)
	B_{j;k+1,k+2} \gamma_{\infty,s}^{(k+2)}  
\eeqn
where
\eqn
	\cU^{(k)}(t) \, \gamma_{\infty,s}^{(k)} \, := \, e^{it  \Delta_{\pm}^{(k)} } \, 
	\gamma_{\infty,s}^{(k)} \,,
\eeqn
and  
\eqn
	\Delta_{\pm}^{(k)} \, = \, \Delta_{\ux_k} - \Delta_{\ux'_k}
\eeqn
with
\eqn
	\Delta_{\pm, x_j} \, = \, \Delta_{x_j} - \Delta_{x_j'} 
	\; \; , \; \; \; 
	\Delta_{\ux_k} \, = \, \sum_{j=1}^k\Delta_{x_j} \,.
\eeqn
Accordingly, it is easy to see that 
\eqn
	\gamma_{\infty,t}^{(k)} \, = \, \big| \, \phi_t \, \ket \bra \, \phi_t \, \big|^{\otimes k}
\eeqn
is a solution of (\ref{eq-BBGKY-intform-1}) if $\phi_t$ satisfies the quintic NLS
\eqn
	i\partial_t \phi_t \, + \, \Delta_x \phi_t \, - \, b_0 \, |\phi_t|^4 \, \phi_t \, = \, 0
\eeqn
with $\phi_0\in L^2(\R^d)$.

In particular, the above described formalism can be rigorously justified, which is the main 
goal of this paper. More precisely, we prove the following theorem in the work at hand:

\begin{theorem}
\label{thm-main-1}
Assume that $d\in\{1,2\}$, and that $V\in W^{2,p}$ for $p\geq 4$ if $d=2$, and
$p>1$ if $d=1$. Moreover, assume that $V(x,x')=V(x',x)$, $V\geq0$, and $0<\beta<\frac1{4(d+1)}$. 
Let $\{\Psi_N\}_N$ denote a family such that 
$$\sup_{N}\frac{1}{N}\bra\Psi_N,\HN\Psi_N\ket<\infty,$$
and which exhibits asymptotic factorization; that is, there exists $\phi\in L^2(\R^d)$ such
that $\tr\big| \, \gamma_N^{(1)}- |\phi \rangle \langle \phi | \, \big|\rightarrow0$ as $N\rightarrow\infty$.
Then, it follows for the $k$-particle marginals $\gamma_{N,t}^{(k)}$ associated
to $\Psi_{N,t}=e^{-it \HN}\Psi_N$ that
\eqn
	\tr\Big| \, \gamma_{N,t}^{(k)} \, - \,  
	| \, \phi_t \rangle \langle \phi_t \, |^{\otimes k}\, \Big|
	\, \rightarrow \, 0
	\; \; \; \; \; \; (N\rightarrow\infty)
\eeqn
where $\phi_t$ solves the defocusing quintic nonlinear Schr\"odinger equation
\eqn\label{eq-quinticNLS-1}
	i\partial_t\phi_t \, + \, \Delta \phi_t \, - \, b_0 \, |\phi_t|^4 \phi_t \, = \, 0 \,
\eeqn
with initial condition $\phi_0=\phi$, and with $b_0=\int_{\R^{2d}} dx \, dx' \, V(x,x')$. 
\end{theorem}

Before we describe the main ideas used in the proof of Theorem \ref{thm-main-1}, 
we give a brief summary of related results.

The mathematical study of systems of interacting Bose gases is a central research area
in mathematical physics which has in recent years experienced remarkable progress. 
A problem of fundamental importance is to prove, in mathematically rigorous terms,
that Bose-Einstein condensation occurs for such systems.
Fundamental progress in the understanding of this problem
and its solution in crucial cases, 
is achieved by Lieb, Seiringer, Yngvason,
et al., in a landmark body of work; see for instance \cite{ailisesoyn,lise,lisesoyn,liseyn}
and the references therein.

A related fundamental line of research addresses the derivation of the mean field
dynamics for a dilute Bose gas, in a scaling regime where the interparticle interactions
and the kinetic energy are comparable in magnitude (the {\em Gross-Pitaevskii scaling}).
Some important early results were obtained in \cite{he,sp}. 
In a highly influential series of works, Erd\"os, Schlein and Yau have proved
for a Bose gas in $\R^3$, with a pair interaction potential that scales to a delta distribution
for particle number $N\rightarrow\infty$, that the limiting dynamics
is governed by a cubic NLS, see \cite{eesy,esy,ey,sc} and the references therein. 
In their approach, the BBGKY hierarchy of $k$-particle marginal density matrices is proven
to converge to an infinite limiting hierarchy (the {\em Gross-Pitaevskii (GP) hierarchy})
in the limit $N\rightarrow\infty$, and the existence and uniqueness of solutions
is established for the infinite hierarchy.
Their uniqueness proof uses sophisticated Feynman 
diagram expansions which are closely related to renormalization methods in quantum field theory,
and represents the most involved part of their analysis.

Recently, Klainerman and Machedon \cite{klma} have developed  a method to prove the 
uniqueness of solutions of the GP hierarchy in $\R^3$ in a class of density matrices 
which differs from the one in \cite{esy}, defined by the 
assumption of certain a priori space-time bounds. 
Subsequently, Kirkpatrick, Schlein and Staffilani have verified a variant
of those a priori bounds
for the model on $\R^2$ and on the torus $\Tor^2$, and derived
the corresponding mean-field limits, \cite{kiscst}.
In dimensions $d \geq 3$ it is currently not known whether 
the limit points obtained from the BBGKY hierarchy in the weak,
subsequential limit $N\rightarrow\infty$ indeed
satisfy this space-time a priori bound. 
For a rigorous derivation of the cubic NLS in dimension $1$,
we refer to \cite{adbagote, adgote}.

Control of the rate of convergence of the quantum evolution 
towards a mean-field limit of Hartree type as $N\rightarrow\infty$
has recently been obtained by Rodnianski and Schlein, \cite{rosc}; 
see also \cite{grmama}. 
The derivation of mean-field limits based on operator-theoretic methods
is developed in work of Fr\"ohlich et al., \cite{frgrsc,frknpi,frknsc};
see also \cite{ansi}.

All of the works cited above investigate properties of Bose gases with pair interactions.
However, in many situations, more general interactions are of importance.
For instance, if the Bose gas interacts with a background
field of matter (such as phonons or photons),
averaging over the latter will typically lead to a linear combination of effective
(renormalized, in the sense of quantum field theory) $n$-particle 
interactions, $n=2,3,\dots$.
For systems exhibiting effective interactions of this general structure, 
it remains a key problem to determine the mean field dynamics.
For $n$-particle interactions with $n=2,3$, where the microscopic Hamiltonian would have
the form
\eqn\label{eq-def-Ham-0-rev}
	\HN & := & \sum_{j=1}^N (-\Delta_{x_j}) \, + \, \frac{1}{N }
	\sum_{1\leq i < j \leq N} N^{d\beta} V_2(N^\beta(x_i-x_j) )
	\\
	&&\quad\quad\quad\quad\quad
	\, + \, \frac{1}{N^2}
	\sum_{1\leq i < j < k \leq N} N^{2d\beta} V_3(N^\beta(x_i-x_j),N^\beta(x_i-x_k)) \,,
	\nonumber
\eeqn
a combination of the analysis given in \cite{kiscst} with the 
one presented here will straightforwardly produce 
a mean field limit described by the defocusing NLS 
\eqn\label{eq-NLS-mixed-1}
	i\partial_t\phi_t+\Delta\phi_t-\lambda_2|\phi_t|^2\phi_t-\lambda_3|\phi_t|^4\phi_t \, = \, 0
\eeqn
in $d=1,2$, where $\lambda_2=\int dx V_2(x)\geq0$
and $\lambda_3=\int dx dx' V_3(x,x')\geq0$ account for the mean-field strength of
the $2$- and $3$-body interactions.  

We shall next outline the approach pursued in this paper. 
We prove Theorem \ref{thm-main-1}  by modifying strategies developed by 
Erd\"os-Schlein-Yau \cite{esy}, Klainerman-Machedon \cite{klma} and
Kirkpatrick-Schlein-Staffilani \cite{kiscst}. More precisely, we prove 
the convergence of the BBGKY hierarchy to the GP hierarchy by   
a straightforward adaptation of the arguments developed in the work \cite{esy}  
(the details are given in
sections \ref{sect-apriori-1} and \ref{sect-compactness-1}; here, we follow the exposition 
of these arguments as presented in \cite{kiscst}).  In order to prove 
the uniqueness of the limiting hierarchy, we expand the method via spacetime norms
introduced 
in \cite{klma}, and subsequently employed in \cite{kiscst}. Roughly speaking, this 
approach consists of the following two main ingredients: 
\begin{enumerate} 
\item First, one expresses the solution $\gamma^{(k)}$ associated to the infinite hierarchy 
\eqref{eq-def-b0-1} in terms of   the subsequent terms  
$\gamma^{(k+2)}$, ... , $\gamma^{(k+2n)}$ by iterating Duhamel's formula.
A key difficulty in controlling the resulting expansion stems from the fact that 
the second expression on the rhs of \eqref{eq-def-b0-1} involves a sum of $k$ terms, so that 
iterating Duhamel's formula produces $k(k+2)...(k+2n-2)$ terms, which is too large
to allow for termwise norm estimates combined with resumming. 
Reformulating a method based on the regrouping of Feynman graphs into equivalence classes 
that was introduced by Erd\"os, Schlein and Yau in \cite{esy}, 
Klainerman and Machedon present an elegant  ``board game'' strategy
to regroup the Duhamel expansion into much fewer $O(C^n)$ sets of terms, \cite{klma};
this allows one to keep track of all
the relevant combinatorics. 
Inspired by \cite{klma}, we define a different board game 
adapted to the new operators $B_{j;k+1,k+2}$ appearing in our limiting hierarchy. 
This new board game allows us to organize the Duhamel expansions in a similar manner as 
in \cite{klma}. 
 
\item Establishing two types of bounds: 
\begin{enumerate} 
\item Space-time $L^2_t L^2_x$ bounds for the freely evolving limiting hierarchy 
(see Theorem \ref{thm-spacetime-bd-1}),
which are used recursively along the iterated Duhamel expansions.
\item Spatial a priori $L^2_x$ bounds for the  limiting hierarchy 
(see Theorem \ref{KMbound}).
\end{enumerate}
In the case $d = 2$, 
we prove both types of bounds, in a similar way as the authors of \cite{kiscst}  in the context of the 2-body 
limiting hierarchy. On the other hand, when $d=1$, the argument used to produce 
$L^2_t L^2_x$ bound of the type (a) for the freely evolving limiting hierarchy 
would produce a divergent bound;
instead, we establish a different  spatial bound (stated in Theorem \ref{highregbound})
for the full limiting hierarchy.  We use this bound iteratively, and at the end combine it  with 
the spatial bound of the type (b). 
\end{enumerate}

\subsection*{Organization of the paper} 
In section \ref{sect-apriori-1} we derive a-priori energy bounds for solutions 
to the BBGKY hierarchy. In section \ref{sect-compactness-1} we summarize main steps 
in the proof of compactness of the sequence of $k$-particle marginals and their convergence 
to the infinite hierarchy. In Section \ref{sect-boundsGP-1} 
we present three types of spatial bounds on the limiting hierarchy, while 
in section \ref{sect-boundsfreeGP-1} we give spacetime bounds on the 
freely evolving infinite hierarchy. 
The sections \ref{sect-uniq-1} - \ref{sect-uniqproof-1}
are devoted to the proof of uniqueness of the limiting hierarchy. 
In particular, in section \ref{sect-uniq-1} 
we state the theorem that guaranties uniqueness
of the infinite hierarchy, while section \ref{sect-Duhamel-1} 
concentrates on combinatorial arguments 
that will be used (together with results 
of sections \ref{sect-boundsGP-1} and \ref{sect-boundsfreeGP-1})
in section \ref{sect-uniqproof-1}, where the uniqueness result is proved.

\section{A priori energy bounds} \label{sect-apriori-1}

To begin with, we derive a priori bounds of the form 
\eqn
	\tr(1-\Delta_{x_1})\cdots(1-\Delta_{x_k}) \gamma_{N,t}^{(k)} 
	\, < \, C^k
\eeqn
which are obtained from energy conservation, following \cite{ey,esy} and \cite{kiscst}.

\begin{proposition}
There exists a constant $C$, and for every $k$, there exists $N_0(k)$ such that
for all $N\geq N_0(k)$, 
\eqn \label{apriori-1-boundNenergy}
	\bra \, \psi \, , \, (\HN+N)^k \, \psi \, \ket \, \geq \, C^k N^k \bra \, \psi \, , 
	\, (1-\Delta_{x_1}) \, \cdots \, (1-\Delta_{x_k}) \, \psi \, \ket 
\eeqn
for all $\psi\in L_s^2(\R^{dN})$.
\end{proposition}

\prf
We adapt the proof in \cite{kiscst} to the current case, which is based on induction in $k$.
We first note that for $k=0$, the statement is trivial, and that for $k=1$, it follows
from $\VN\geq0$.
For the induction step, we assume that for all $k\leq n$, the statement is correct. 
We then prove its validity for $n+2$. Following \cite{kiscst}, 
we write $S_i=(1-\Delta_{x_i})^{1/2}$ and
$\HN+N=h_1+h_2$ with
\eqn
	h_1 & =& \sum_{j=n+1}^N S_j^2
	\nonumber\\
	h_2 & = & \sum_{j=1}^n S_j^2 \, + \, \sum_{1\leq i < j < \ell \leq N}
	N^{-2}  \VN( x_i-x_j , x_i-x_\ell ) \,.
\eeqn
Using the induction assumption, we infer that
\eqn
	\lefteqn{
	\bra \, \psi \, , \, (\HN+N)^{n+2} \, \psi \, \ket \, 
	}
	\nonumber\\
	& \geq & 
	C^n N^n \bra \, \psi \, , \, (\HN+N) \, S_1^2 \cdots \, S_n^2 \, (\HN+N) \, \psi \, \ket 
	\nonumber\\
	& \geq & 
	C^n N^n \bra \, \psi \, , \, h_1 \, S_1^2 \cdots \, S_n^2 \, h_1 \, \psi \, \ket 
	\, + \, 
	C^n N^n \big( \,
	\bra \, \psi \, , \, h_1 \, S_1^2 \cdots \, S_n^2 \, h_2 \, \psi \, \ket \, + \, c.c. \, \big) 
	\nonumber\\
	&\geq&
	C^n N^n (N-n)(N-n-1)\bra \, \psi \, , \, S_1^2 \cdots \, S_{n+2}^2 \, \psi \, \ket 
	\nonumber\\
	&&
	\, + \, 
	C^n N^n (N-n) n \bra \, \psi \, , \, S_1^4 S_2^2\cdots \, S_{n+1}^2 \, \psi \, \ket 
	\nonumber\\
	&&
	\, + \,
	C^n N^n \frac{(N-n) }{N^2} \cdot
	\\
	&&
	\cdot\sum_{1\leq i<j<\ell\leq N} 
	\big( 
	\bra \, \psi \, , \, S_1^2 \cdots \, S_{n+1}^2 
	\, \VN(x_i-x_j,x_i-x_\ell) \, \psi \, \ket 
	\, + \, c.c. \, \big) \,.
	\nonumber
\eeqn 
Due to the permutation symmetry of $\psi$ in all of its arguments, 
it follows that there exists $N_0(n)$ such that for all $N>N_0(n)$,
\eqn\label{eq-energybd-aux-1}
	\lefteqn{
	\bra \, \psi \, , \, (\HN+N)^{n+2} \, \psi \, \ket \, 
	}
	\nonumber\\
	& \geq & 
	C^{n+2} N^{n+2} \bra \, \psi \, , \, S_1^2 \cdots \, S_{n+2}^2 \, \psi \, \ket 
	\, + \, 
	C^{n+1} N^{n+1} \bra \, \psi \, , \, S_1^4 S_2^2\cdots \, S_{n+1}^2 \, \psi \, \ket 
	\label{eq-enaprbd-aux-0}\\
	&&
	+\Big[ C^n N^{n-2} (N-n) (N-n-1) (N-n-2) (N-n-3)
	\label{eq-enaprbd-aux-1}\\
	&&\quad\quad\quad\quad\quad\quad\quad\quad\quad\quad
	\bra \, \psi \, , \, S_1^2 \cdots \, S_{n+1}^2 
	\, \VN(x_{n+2}-x_{n+3},x_{n+2}-x_{n+4}) \, \psi \, \ket
	\nonumber\\
	&&
	\, + \, C^n N^{n-2} (N-n) (N-n-1) (N-n-2) (n+1)
	\label{eq-enaprbd-aux-2}\\
	&&\quad\quad\quad\quad\quad\quad\quad\quad\quad\quad
	\bra \, \psi \, , \, S_1^2 \cdots \, S_{n+1}^2 
	\, \VN(x_{1}-x_{n+2},x_{1}-x_{n+3}) \, \psi \, \ket
	\nonumber\\
	&&
	\, + \, C^n N^{n-2} (N-n) (N-n-1) n (n-1)
	\label{eq-enaprbd-aux-3}\\
	&&\quad\quad\quad\quad\quad\quad\quad\quad\quad\quad
	\bra \, \psi \, , \, S_1^2 \cdots \, S_{n+1}^2 
	\, \VN(x_{1}-x_{2},x_{1}-x_{n+2}) \, \psi \, \ket
	\nonumber\\
	&&
	\, + \, C^n N^{n-2} (N-n) (n+1) n (n-1) \bra \, \psi \, , \, S_1^2 \cdots \, S_{n+1}^2 
	\, \VN(x_{1}-x_{2},x_{1}-x_{3}) \, \psi \, \ket
	\nonumber\\
	&& \, + \, c.c. \, \Big] \,.
	\label{eq-enaprbd-aux-4}
\eeqn
We will now discuss individually each of the terms in  
\eqref{eq-enaprbd-aux-1} - \eqref{eq-enaprbd-aux-4}.
For the term \eqref{eq-enaprbd-aux-1}, we note that  all $S_1,\dots,S_{n+1}$
commute with $V_N(x_{n+2}-x_{n+3},x_{n+2}-x_{n+4})$. Therefore, we have
\eqn
	\lefteqn{
	\bra \, \psi \, , \, S_1^2 \cdots \, S_{n+1}^2 
	\, \VN(x_{n+2}-x_{n+3},x_{n+2}-x_{n+4}) \, \psi \, \ket
	}
	\\
	&& = \, \int d\ux_N \VN(x_{n+2}-x_{n+3},x_{n+2}-x_{n+4})
	\, |(S_1 \cdots \, S_{n+1} \psi)(\ux_N) |^2 \, \geq \, 0
	\nonumber  
\eeqn
which is positive, due to the positivity of $\VN$.
Thus, for a lower bound on (\ref{eq-energybd-aux-1}) --  (\ref{eq-enaprbd-aux-4}), the term
 \eqref{eq-enaprbd-aux-1} can be discarded.


For  \eqref{eq-enaprbd-aux-2}, we let $\dot S_j=(S_{j,i})_{i=1}^d:=i\nabla_{x_j}$
so that $S_j^2=1+\dot S_j^2$. Moreover, we will in the sequel use the notation
$\dot S_j \cdots \dot S_j$ for $\sum_{i=1}^d \dot S_{j,i} \cdots \dot S_{j,i}$ 
(sum over coordinate components for  pairs of $\dot S_j$'s with the same index $j$). 
Then, 
\eqn
	\lefteqn{
	\bra \, \psi \, , \, S_1^2 \cdots \, S_{n+1}^2 
	\, \VN(x_{1}-x_{n+2},x_{1}-x_{n+3}) \, \psi \, \ket
	}
	\nonumber\\
	&\geq&  
	\bra \, \psi \, , \, S_{n+1}\cdots S_2  
	\VN(x_{1}-x_{n+2},x_{1}-x_{n+3}) S_2\cdots \, S_{n+1} \, \psi \, \ket 
	\nonumber\\
	&& + 
	\, \bra \, \psi \, , \, S_{n+1}\cdots S_2 \dot S_1
	[\dot S_1,\VN(x_{1}-x_{n+2},x_{1}-x_{n+3})] S_2\cdots \, S_{n+1} \, \psi \, \ket|
	\nonumber\\
	&\geq&  \label{cdots-term2-1-1}
	- \, |\bra \, \psi \, , \, S_{n+1}\cdots S_2 \dot S_1
	[\dot S_1,\VN(x_{1}-x_{n+2},x_{1}-x_{n+3})] S_2\cdots \, S_{n+1} \, \psi \, \ket|
	\\
	&=&-  \, |\bra \, \psi \, , \, S_{n+1} \cdots S_2\dot  S_1  (\nabla_{x_1}\VN(x_{1}-x_{n+2},x_{1}-x_{n+3}))  
	S_2\cdots \, S_{n+1} \, \psi \, \ket|
	\nonumber\\
	&\geq&- \mu \, |\bra \, \psi \, , \, S_{n+1}^2 \cdots S_1^2 \, \psi \, \ket|
	\nonumber\\
	&&- \mu^{-1} \, |\bra \, \psi \, , \, S_{n+1} \cdots S_2 
	| \nabla_{x_1}\VN(x_{1}-x_{n+2},x_{1}-x_{n+3}) | ^2
	S_2\cdots \, S_{n+1} \, \psi \, \ket|
	\nonumber\\
	&\geq&-  \, \mu\, |\bra \, \psi \, , \, S_{n+1}^2 \cdots S_1^2 \, \psi \, \ket|
	\nonumber\\
	&& -C \, \mu^{-1} \, \|\nabla \VN\|_{L^{2p}(\R^{2d})}^2
	\bra \, \psi \, , \,  \, S_1^2 \cdots \, S_{n+2}^2  \, \psi \, \ket \,, \label{cdots-term2}
\eeqn 
for $p \geq 4$ if $d=2$, and $p>1$ if $d=1$, where we used the Schwarz inequality and 
Lemma \ref{lm-VSob-bd-1}.
We remark that in the pass to \eqref{cdots-term2-1-1}, we have used the positivity of $\VN\geq0$.
As a consequence of 
\eqn \label{apriori-nabla-scal}
	\|\nabla \VN\|_{L^{2p}(\R^{2d})} \, \leq \, C N^{2\beta(d+\frac12-\frac d{2p})}
	\|\nabla V\|_{L^{2p}(\R^{2d})} \,,
\eeqn
the expression \eqref{cdots-term2} implies 
\eqn
	\lefteqn{
	\bra \, \psi \, , \, S_1^2 \cdots \, S_{n+1}^2 
	\, \VN(x_{1}-x_{n+2},x_{1}-x_{n+3}) \, \psi \, \ket
	}
	\nonumber\\
	&& 
	\quad\quad\quad \geq \,
	- \, C \, N^{2\beta(2d+1)} \, \bra \, \psi \, , \, S_1^2 \cdots \, S_{n+2}^2 \, \psi \, \ket
\eeqn
for a constant $C>0$. 

For the term  \eqref{eq-enaprbd-aux-3}, we expand 
$S_1^2S_2^2=(1+\dot S_1^2)(1+\dot S_2^2)$, and get
\eqn
	\lefteqn{
	\bra \, \psi \, , \, S_1^2 \cdots \, S_{n+1}^2 
	\, \VN(x_{1}-x_{2},x_{1}-x_{n+2}) \, \psi \, \ket
	}
	\nonumber\\
	&=&  \bra \, \psi \, , \,   S_{n+1} \cdots S_3
	\, \VN(x_{1}-x_{2},x_{1}-x_{n+2})  \, S_3 \cdots \, S_{n+1}\, \psi \, \ket
	\nonumber\\
	&&+ \,
	 \bra \, \psi \, , \, (\dot S_1^2+\dot S_2^2)S_3^2 \cdots \, S_{n+1}^2 
	\, \VN(x_{1}-x_{2},x_{1}-x_{n+2}) \, \psi \, \ket
	\nonumber\\
	&&+ \,
	 \bra \, \psi \, , \, \dot S_1^2 \dot S_2^2 S_3^2 \cdots \, S_{n+1}^2 
	\, \VN(x_{1}-x_{2},x_{1}-x_{n+2}) \, \psi \, \ket
	\nonumber\\
	&\geq&  
	\sum_{j=1,2} \bra \, \psi \, , \,  S_{n+1} \cdots S_3
	\, \dot S_j [\dot S_j,\VN(x_{1}-x_{2},x_{1}-x_{n+2})] \, S_3 \cdots \, S_{n+1} \psi \, \ket
	\\
	&&+\,
	  \bra \, \psi \, , \,  S_{n+1} \cdots S_3
	\, \dot S_2 \dot S_1 [\dot S_1 \dot S_2  ,\VN(x_{1}-x_{2},x_{1}-x_{n+2})] 
	\, S_3 \cdots \, S_{n+1} \psi \, \ket \,.
	\nonumber
\eeqn
Both terms in the sum on the second last line can be treated as in \eqref{cdots-term2}, and obey the
same lower bound \eqref{cdots-term2}.
For the term on the last line, we use that
\eqn
	 [\dot S_1 \dot S_2  ,\VN(\cdots)] \, = \, 
	 [\dot S_1  ,\VN(\cdots)] \dot S_2   \, + \, \dot S_1 [\dot S_2  ,\VN(\cdots)]\,.
\eeqn
Accordingly,
\eqn 
	\bra \, \psi \, , \,  S_{n+1} \cdots S_3
	\, \dot S_1 \dot S_2 [\dot S_1 \dot S_2  ,\VN(x_{1}-x_{2},x_{1}-x_{n+2})] 
	\, S_3 \cdots \, S_{n+1} \psi \, \ket
	\, \geq \,  (I) \, + \, (II)
	\nonumber\\
\eeqn
where
\eqn
	(I)&:=&- \, |\bra \, \psi \, , \, S_{n+1} \cdots S_3 \dot S_2 \dot S_1^2
	[\dot S_2,\VN(x_{1}-x_{2},x_{1}-x_{n+2})] S_3\cdots \, S_{n+1} \, \psi \, \ket|
	\nonumber\\
	(II)&:=&- \, |\bra \, \psi \, , \, S_{n+1} \cdots S_3 \dot S_2 \dot S_1
	[\dot S_1,\VN(x_{1}-x_{2},x_{1}-x_{n+2})]\dot S_2  S_3\cdots \, S_{n+1} \, \psi \, \ket| \,.
	\nonumber\\
\eeqn
We obtain 
\eqn
	(I)&\geq& - 2 \nu \, |\bra \, \psi \, , \, S_{n+1}^2 \cdots S_2^2 S_1^4 \, \psi \, \ket|
	\nonumber\\
	&&- \,
	\nu^{-1} \, |\bra \, \psi \, , \, S_{n+1} \cdots S_3  
	 | \nabla_{x_2}  \VN(x_{1}-x_{2},x_{1}-x_{n+2}) | ^2
	S_3\cdots \, S_{n+1} \, \psi \, \ket|
	\nonumber\\ 
        &\geq&-  \, \nu\, |\bra \, \psi \, , \, S_{n+1}^2 \cdots S_2^2 S_1^4 \, \psi \, \ket|
	\nonumber\\
	&& -C \, \nu^{-1} \, \|  \VN\|^2_{W^{1,2p}(\R^{2d})}
	\bra \, \psi \, , \,  \, S_1^2 \cdots \, S_{n+1}^2  \, \psi \, \ket
	\nonumber \\  
        & \geq & -  \, \nu\, |\bra \, \psi \, , \, S_{n+1}^2 \cdots S_2^2 S_1^4 \, \psi \, \ket|
	\nonumber\\
	&&-C \; \nu^{-1} N^{4\beta(d+\frac{1}{2}-\frac{d}{2p})} \|V\|^2_{W^{1,2p}(\R^{2d})}
	\bra \, \psi \, , \,  \, S_1^2 \cdots \, S_{n+1}^2  \, \psi \, \ket   \,, \label{cdots-term3}
\eeqn
where we used the Schwarz inequality, Lemma \ref{lm-VSob-bd-1} and \eqref{apriori-nabla-scal}.
Moreover,  similarly to \eqref{cdots-term2}, we obtain
\eqn 
	(II) &\geq&-  \, \mu\, |\bra \, \psi \, , \, S_{n+1}^2 \cdots S_2^2 S_1^2 \, \psi \, \ket|
	\label{cdots-term2-2}\\
	&& -C \, \mu^{-1} \, \|\nabla \VN\|_{L^{2p}(\R^{2d})}^2
	\bra \, \psi \, , \,  \, S_1^2 S_2^2 \cdots \, S_{n+2}^2  \, \psi \, \ket   \,,
	\nonumber
\eeqn
using  Lemma \ref{lm-VSob-bd-1}.
 
Finally, for term \eqref{eq-enaprbd-aux-4}, we use
\eqn 
	\bra \, \psi \, , \, S_1^2 \cdots \, S_{n+1}^2 
	\, \VN(x_{1}-x_{2},x_{1}-x_{3}) \, \psi \, \ket 
	\, = \,  (A_0) \, + \, (A_1) \, + \, (A_2) \, + \, (A_3) \, 
\eeqn
where
\eqn
	(A_0)
	&:=& \bra \, \psi \, , \,   S_{n+1} \cdots S_4
	\, \VN(x_{1}-x_{2},x_{1}-x_{3})  \, S_4 \cdots \, S_{n+1}\, \psi \, \ket \, \geq \, 0
	\nonumber\\
	(A_1)&:=& \bra \, \psi \, , \, (\dot S_1^2+\dot S_2^2 + \dot S_3^2) S_4^2 \cdots \, S_{n+1}^2 
	\, \VN(x_{1}-x_{2},x_{1}-x_{3}) \, \psi \, \ket
	\nonumber\\
	(A_2)&:=& \sum_{1\leq i<j\leq3}
	 \bra \, \psi \, , \, (\dot S_i^2\dot S_j^2) S_4^2 \cdots \, S_{n+1}^2 
	\, \VN(x_{1}-x_{2},x_{1}-x_{3}) \, \psi \, \ket
	\nonumber\\
	(A_3) &:=&  
	 \bra \, \psi \, , \, \dot S_1^2\dot S_2^2 \dot S_3^2 S_4^2 \cdots \, S_{n+1}^2 
	\, \VN(x_{1}-x_{2},x_{1}-x_{3}) \, \psi \, \ket \, \,.
\eeqn
The term $(A_0)$ is positive, and can be discarded for a lower bound.
The term $(A_1)$ can be bounded as in  \eqref{cdots-term2},
and the term $(A_2)$ as in  \eqref{cdots-term3}.
We are left with
\eqn
	(A_3)
	\, = \, \bra \, \psi \, , \, S_{n+1} \cdots S_4 \dot S_3 \dot S_2 \dot S_1 
	[\dot S_1 \dot S_2 \dot S_3,\VN(x_{1}-x_{2},x_{1}-x_{3})] S_4\cdots \, S_{n+1} \, \psi \, \ket 
	\quad
\eeqn
where we note that
\eqn 
	\lefteqn{ 
	 [\dot S_1 \dot S_2 \dot S_3 ,\VN(\cdots)]
	}
	\nonumber\\
	& = & 
	 [\dot S_1  ,\VN(\cdots)] \dot S_2  \dot S_3  \, + \, \dot S_1 [\dot S_2  ,\VN(\cdots)] \dot S_3 
	\, + \, \dot S_1 \dot S_2 [\dot S_3  ,\VN(\cdots)]  \,.
	\nonumber\\
\eeqn
Accordingly,
\eqn
	(A_3) \, = \,(A_3)_1 \, + \, (A_3)_2 \,+ \, (A_3)_3
\eeqn
where
\eqn
	(A_3)_1 \, := \,  \bra \, \psi \, , \, S_{n+1} \cdots S_4 \dot S_3 \dot S_2 \dot S_1 
	[\dot S_1  ,\VN(\cdots)] \dot S_2  \dot S_3  S_4\cdots \, S_{n+1} \, \psi \, \ket 
	\label{eq-energybd-aux-2-3-1}
\eeqn
can be treated as in   \eqref{cdots-term2-2}, and is bounded from below by   \eqref{cdots-term2-2}.
Moreover,
\eqn 
	(A_3)_2 \, := \,  \bra \, \psi \, , \, S_{n+1} \cdots S_4 \dot S_3 \dot S_2 \dot S_1 
	 \dot S_1 [\dot S_2  ,\VN(\cdots)] \dot S_3    S_4\cdots \, S_{n+1} \, \psi \, \ket 
	\label{eq-energybd-aux-2-3-2}
\eeqn
can be treated as in \eqref{cdots-term3}, and is bounded from below by   \eqref{cdots-term3}.
To bound the term
\eqn  
	(A_3)_3 \, := \,  \bra \, \psi \, , \, S_{n+1} \cdots S_4 \dot S_3 \dot S_2 \dot S_1 \,
	 ( \dot S_1 \dot S_2 [\dot S_3  ,\VN(\cdots)] ) \, S_4\cdots \, S_{n+1} \, \psi \, \ket 
	\label{eq-energybd-aux-2-3-3}
\eeqn 
from below, we cannot proceed directly as in \eqref{cdots-term3} because we would thereby
obtain a term of the form $- |\bra \, \psi \, , \, S_{n+1}^2 \cdots S_3^2 S_2^4 S_1^4 \, \psi \, \ket|$
which we do not know how to control. We instead use that
\eqn 
	 \dot S_1 \dot S_2 [\dot S_3  ,\VN(\cdots)]  & = & 
	\dot S_1[\dot S_3   , ( \dot S_2\VN(\cdots))] \, + \, 
	\dot S_1[\dot S_3   , \VN(\cdots))]  \dot S_2 
	\nonumber\\
	 & = & 
	-\dot S_1  (\nabla_{x_2} \nabla_{x_3}\VN(\cdots)) \, + \, 
	\dot S_1(i\nabla_{x_3}\VN(\cdots))]  \dot S_2 \,,
	\quad
\eeqn
which follows from the fact that the operators $\dot S_j$ are derivations and satisfy the 
Leibnitz rule. Accordingly, 
\eqn 
	(A_3)_3 \, := \, (A_3)_{3,1} \, + \, (A_3)_{3,2} 
\eeqn
where
\eqn 
	(A_3)_{3,1} \, := \, -\, \bra \, \psi \, , \, S_{n+1} \cdots S_4 \dot S_3 \dot S_2 
	\dot S_1^2  (\nabla_{x_2} \nabla_{x_3}\VN(\cdots)) S_4\cdots \, S_{n+1} \, \psi \, \ket 
	\label{eq-energybd-aux-2-3-3-1}
\eeqn
and
\eqn 
	(A_3)_{3,2} \, := \, -\, \bra \, \psi \, , \, S_{n+1} \cdots S_4 \dot S_3 \dot S_2  
	\dot S_1^2 (i\nabla_{x_3}\VN(\cdots))  \dot S_2  S_4\cdots \, S_{n+1} \, \psi \, \ket \,.
	\label{eq-energybd-aux-2-3-3-2}
\eeqn
Both terms can be treated as in  \eqref{cdots-term3}, and are  bounded from below by the 
quantity obtained from replacing $\VN$ in the lower bound \eqref{cdots-term3} by $|\nabla \VN|$.
This implies that
\eqn 
	(A_3) & \geq &  -  \, \nu\, |\bra \, \psi \, , \, S_{n+1}^2 \cdots S_2^2 S_1^4 \, \psi \, \ket|
	\nonumber\\
	&&-C \; \nu^{-1} N^{4\beta(d+1-\frac{d}{2p})} \|V\|^2_{W^{2,2p}(\R^{2d})}
	\bra \, \psi \, , \,  \, S_1^2 \cdots \, S_{n+1}^2  \, \psi \, \ket  
\eeqn
as one easily verifies by comparing with  \eqref{cdots-term3}.

In conclusion, the sum of terms inside  $[\cdots]$ in  \eqref{eq-enaprbd-aux-1} - \eqref{eq-enaprbd-aux-4}
is bounded from below by
\eqn
	[\cdots] & \geq &
	-  \,C(n) N^{n} |\bra \, \psi \, , \, S_{n+1}^2 \cdots S_2^2 S_1^4 \, \psi \, \ket|
	\nonumber\\
	&&- \, C(n,V) N^{n-1+4\beta(d+1)} 
	\bra \, \psi \, , \,  \, S_1^4 S_2^2 \cdots \, S_{n+1}^2  \, \psi \, \ket \,,
\eeqn
which is dominated by the first two  terms on the rhs of  (\ref{eq-energybd-aux-1})
(both of which are positive), 
for $\beta<\frac1{4(d+1)}$ and  large $N$. 
This immediately establishes the induction step $n\rightarrow n+2$.
\endprf

\begin{lemma}\label{lm-VSob-bd-1}
For dimensions $d\geq1$, the estimate
\eqn
	\lefteqn{
	\bra \, \psi_1 \, , \, V(x_1,x_2) \, \psi_2 \, \ket 
	}
	\\
	& \leq &
	C_{p,d} \, \|V\|_{L^p_{x_1,x_2}} 
	\| \,    \psi_1 \, \|_{L^2_{x_1,x_2}} \, 
	\| \, \langle\nabla_{x_1}\rangle\langle\nabla_{x_2}\rangle \,  \psi_2 \, \|_{L^2_{x_1,x_2}} 
	\nonumber
\eeqn
holds for any $p\geq 2d$ if $d\geq2$, and for any $p>1$ if $d=1$.
\end{lemma}

\prf
Clearly, using the H\"older (for $1=\frac1p+\frac1{2}+\frac1{q}$) and Sobolev inequalities,
\eqn
	| \, \bra \, \psi_1 \, , \, V(x_1,x_2) \, \psi_2 \, \ket \, |  
	&\leq&
	C_{p,d} \,\|V \|_{L^p_{x_1,x_2}} \| \psi_1 \|_{L^{2 }_{x_1,x_2}} \| \psi_2 \|_{L^{q}_{x_1,x_2}} 
	\nonumber\\ 
	& \leq & C_{p,d} \,  \|V\|_{L^{p}_{x_1,x_2}} \| \psi_1 \|_{L^{2}_{x_1,x_2}} \| \psi_2 \|_{H^1_{x_1,x_2}}
\eeqn
provided that $2\leq q\leq \frac{4d}{2d-2}$ if $d\geq2$ (interpreting $(x_1,x_2)$ as a point in $\R^{2d}$),
and $2\leq  q<\infty$ if $d=1$.
This immediately implies that $d \leq p <\infty$ for $d\geq2$, 
and  $1 < p <\infty$ for $d=1$.
Moreover, it is clear that
\eqn
	 \| \psi \|^2_{H^1_{x_1,x_2}} & = & 
	 \bra \, \psi \, , \, (1-\Delta_{x_1} -\Delta_{x_2}) \, \psi \, \ket
	\nonumber\\ 
	& \leq & \bra \, \psi \, , \, (1-\Delta_{x_1})(1-\Delta_{x_2}) \, \psi \, \ket \,,
\eeqn
from $\bra \, \psi \, , \,  \Delta_{x_1} \Delta_{x_2} \, \psi \, \ket=\|\nabla_{x_1}\nabla_{x_2}\psi\|^2_{L^2}\geq0$.
The claim follows immediately.
\endprf

In conclusion, we have found the following a priori estimate.

\begin{corollary}
\label{cor-energy-apbd-1}
Define 
\eqn \label{defcutof}
	\tpsi_N \, := \, \frac{\chi(\frac\kappa N \HN) \psi_N}{\|\chi(\frac\kappa N \HN) \psi_N\|}
\eeqn
where $\chi$ is a bump function supported on $[0,1]$, and $\kappa>0$ is a real parameter.
Let $\tpsi_{N,t}=e^{-it\HN}\tpsi_N$, and let $\widetilde\gamma_{N,t}^{(k)}$ be the corresponding
$k$-particle marginal. 
Then, there exists a constant $C$ independent of $k$ and there exists 
an integer $N_0(k)$ for every $k\geq1$, such that for all $N>N_0(k)$,
\eqn \label{bd-trace} 
	\tr(1-\Delta_{x_1})\cdots(1-\Delta_{x_k})\widetilde\gamma_{N,t}^{(k)} \, \leq \, C^k \,.
\eeqn
\end{corollary}

\section{Compactness and convergence to the infinite hierarchy}
\label{sect-compactness-1}

In this section, we summarize
the main steps of the proof in \cite{esy,kiscst}  of the compactness of the 
sequence of $k$-particle marginals and the convergence to the infinite hierarchy
as $N\rightarrow\infty$.
The arguments developed in these works can be adopted almost verbatim for the 
current context.
We outline them here for the convenience of the reader, essentially 
quoting the exposition in \cite{kiscst}.

The following topology   on the space of density matrices is picked in \cite{esy}.
We let $\cK_k=\cK(L^2((\R^d)^k))$ denote the space of compact operators on 
$L^2(\R^d)$ equipped with the
operator norm topology, and $\cL^1_k:=\cL^1(L^2((\R^d)^k))$ denotes the space of trace class operators
on $L^2((\R^d)^k)$ equipped with the trace class norm. It is a standard fact that 
$\cL^1_k=\cK_k^*$. 
$\cK_k$ is a separable Banach space, hence any closed ball in $\cL_k^1$,
which is weak-* compact by the Banach-Alaoglu theorem,
is metrizable in the weak-* topology.
Let $\{J^{(k)}_j\}$ be a countable dense subset 
of the unit ball of $\cK_k$, that is,
$\|J^{(k)}_j\|\leq1$. Then, 
\eqn
	\eta_k(\gamma^{(k)},\widetilde\gamma^{(k)}) \, = \, 
	\sum_{j\in \N} 2^{-j} \Big| \, \tr J^{(k)}_j 
	\big( \, \gamma^{(k)} - \widetilde\gamma^{(k)} \, \big) \, \Big| \,
\eeqn
is a metric, and the associated metric topology is equivalent to the 
weak-* topology. 
A uniformly bounded sequence $\gamma_N^{(k)}\in\cL^1_k$ converges to 
$\gamma^{(k)}\in\cL^1_k$ with respect to the weak-* topology if and only if 
$\eta_k(\gamma^{(k)}_N,\gamma^{(k)})\rightarrow0$ as $N\rightarrow\infty$.

Let $C([0,T],\cL^1_k)$ be the space of $\cL^1_k$-valued functions
of $t\in[0,T]$ that are continuous with respect to the metric $\eta_k$. 
One can endow  $C([0,T],\cL^1_k)$ with the metric 
$\widehat\eta_k(\gamma^{(k)}(\cdot),\widetilde\gamma^{(k)}(\cdot))=
\sup_{t\in[0,T]}\eta_k(\gamma^{(k)}(t),\widetilde\gamma^{(k)}(t))$, \cite{esy}.
This induces the product topology $\tau_{prod}$ on $\oplus_{k\in\N}C([0,T],\cL^1_k)$,
generated by the metrics $\widehat\eta_k$
on $C([0,T],\cL^1_k)$, for $k\in\N$.
For more details, see \cite{esy}.

\begin{proposition}
The sequence of marginal densities 
$\widetilde\Gamma_{N,t}=\{\widetilde\gamma_{N,t}^{(k)}\}_{k=1}^N$
is compact with respect to the product topology 
$\tau_{prod}$ generated by the metrics $\widehat\eta_k$,
\cite{esy}. 
Any subsequential limit point $\Gamma_{\infty,t}=\{ \gamma_{\infty,t}^{(k)}\}_{k\geq1}$ 
has the property
that the components $\gamma_{\infty,t}^{(k)}$ are symmetric under permutations, 
is positive, and $\tr\gamma_{\infty,t}^{(k)}\leq1$
for every $k\geq1$.
\end{proposition}

\prf
The proof is completely analogous to the one given for a related result in \cite{esy},
and for Theorem 4.1 in
\cite{kiscst}. We summarize the main steps.

Using a Cantor diagonal argument, it is sufficient to prove
the compactness of $\widetilde\gamma_{N,t}^{(k)}$
for a fixed $k$. 
By the Arzela-Ascoli theorem,
this is achieved by proving equicontinuity of  
$\Gamma_{N,t}=\{\widetilde\gamma_{N,t}^{(k)}\}_{k=1}^N$ 
with respect to the
metric $\widehat\eta_k$. 
It is sufficient to prove that for every observable $J^{(k)}$ from a dense
subset of $\cK_k$ and for every $\epsilon>0$, there exists $\delta=\delta(J^{(k)},\epsilon)$
such that 
$$
	\sup_{N\geq1}\Big| \, \tr J^{(k)}\big( \, \widetilde\gamma_{N,t}^{(k)} 
	- \widetilde\gamma_{N,s}^{(k)} \, \big) \, \Big| \, < \, \epsilon
$$
for all $t,s\in[0,T]$ with $|t-s|<\delta$.
To this end, the norm 
\eqn \label{triplenorm} 
	|\!|\!| J^{(k)} |\!|\!| \, = \, \sup_{\up_k'}\int d\up_k \prod_{j=1}^k \bra p_j \ket \bra p_j' \ket
	\Big( \, |\widehat J^{(k)}(\up_k;\up_k')| + | \widehat J^{(k)}(\up_k';\up_k)| \, \Big) 
\eeqn
is considered in \cite{esy,kiscst}. Notably, the set of all $J^{(k)}\in\cK_k$
for which this norm is finite, is dense in $\cK_k$.

The claim of the proposition then follows from
\eqn
	\sup_{N\geq1}\Big| \, \tr J^{(k)} \, \big( \, \widetilde\gamma_{N,t}^{(k)} 
	- \widetilde\gamma_{N,s}^{(k)} \, \big) \, \Big| \, < C \, |\!|\!| J^{(k)} |\!|\!| \, |t-s|
\eeqn
which is proved in the same manner as in \cite{esy,kiscst}.
\endprf

\begin{theorem}
Assume that $\Gamma_{\infty,t}=\{\gamma_{\infty,t}^{(k)}\}_{k=1}^\infty\in 
\oplus_{k\geq}C([0,T],\cL^1_k)$
is a limit point of $\widetilde\Gamma_{N,t}=\{\widetilde\gamma_{N,t}^{(k)}\}_{k=1}^N$ 
with respect to the product topology $\tau_{prod}$. Then, $\Gamma_{\infty,t}$
is a solution of the infinite hierarchy
\eqn\label{eq-BBGKY-intform-2} 
	\gamma_{\infty,t}^{(k)} \, = \, \cU^{(k)}(t) \, \gamma_{\infty,0}^{(k)} 
	- \, i \, b_0 \, \sum_{j=1}^k \int_0^t ds \, \cU^{(k)}(t-s) \,
	B_{j;k+1,k+2} \, \gamma_{\infty,s}^{(k+2)}  
\eeqn
with initial data $\gamma_{\infty,0}^{(k)}=\big| \phi\ket\bra\phi\big|^{\otimes k}$.
\end{theorem}

\prf 
Here again, the proof can be adopted in a straightforward
manner from \cite{kiscst}. We outline the 
main steps.

Let us fix $k \geq 1$. By passing to a subsequence we can assume that 
for every $J^{(k)} \in \cK_{k}$  we have: 
\eqn \label{conv-assum} 
                       \sup_{t \in [0,T]} \tr \; J^{(k)} 
                       \left( \widetilde{\gamma}_{N,t}^{(k)} - \gamma_{\infty,t}^{(k)}\right) 
                       \rightarrow 0, \; \; \; \; \mbox{ as } N \rightarrow \infty.
\eeqn
This assumption is based on that fact that $\tr \; J^{(k)}  \left( \widetilde{\gamma}_{N,t}^{(k)} - \gamma_{\infty,t}^{(k)}\right)  \rightarrow 0$ 
as $N \rightarrow \infty$, which was proved by  Erd\"{o}s, Schlein and Yau \
in \cite{ESY-derGP} (see Proposition 8.1), and 
under less restrictive conditions in \cite{ESY-rigderGP} (see Proposition 9.1). 

We shall prove \eqref{eq-BBGKY-intform-2} by testing 
the limit point $\gamma_{\infty,t}^{(k)}$ against an observable belonging to  
a dense set in $\cK_k$. In particular, choose an arbitrary 
$J^{(k)} \in \cK_k$ such that $|\!|\!| J^{(k)} |\!|\!| < \infty$ (see
\eqref{triplenorm} for the definition of
the norm $|\!|\!| \cdot |\!|\!|$).
It suffices to prove that 
\eqn\label{conv-suff1} 
\tr J^{(k)} \gamma_{\infty, 0}^{(k)} = \tr J^{(k)} \big|\phi\ket\bra\phi\big|^{\otimes k}
\eeqn 
and 
\begin{align} 
             \tr \; J^{(k)}\gamma_{\infty,t}^{(k)} 
             & = \tr \; J^{(k)} \cU^{(k)}(t) \, \gamma_{\infty,0}^{(k)} \nonumber \\
	     & - \, i \, b_0 \, \sum_{j=1}^k \int_0^t ds \, \tr \; J^{(k)} 
                 \cU^{(k)}(t-s) \, B_{j;k+1,k+2} \, \gamma_{\infty,s}^{(k+2)}.  
\label{conv-suff2}
\end{align}

First, we note that \eqref{conv-suff1} follows from \eqref{conv-assum}. 
To prove \eqref{conv-suff2}, we write the BBGKY hierarchy \eqref{intro-BBGKY} in integral form,
\begin{align} 
              \tr & \; J^{(k)}\widetilde{\gamma}_{N,t}^{(k)} \label{conv-lhs} \\ 
              & = \tr \; J^{(k)} \cU^{(k)}(t) \, \widetilde\gamma_{N,0}^{(k)} \label{conv-t1} \\
              & - \, \frac{i}{N^2} \, \sum_{1\leq i<j<\ell\leq k} 
                  \int_0^t ds \, \tr \; J^{(k)} \cU^{(k)}(t-s)       
                  [\VN(x_i-x_j , x_i-x_\ell) , \widetilde\gamma_{N,s}^{(k)} ] \label{conv-t2} \\
              & - \; \frac{i(N-k)}{N^2} \; \sum_{1\leq i<j\leq k} 
                  \int_0^t ds \, \tr \; J^{(k)} \cU^{(k)}(t-s) 
                  [\VN(x_i-x_{j}, x_i-x_{k+1}) , \widetilde\gamma_{N,s}^{(k+1)} ] \label{conv-t3} \\
              & - \; \frac{i(N-k)(N-k-1)}{N^2} \; \sum_{j=1}^k
                  \int_0^t ds \, \tr \; J^{(k)} \cU^{(k)}(t-s)  \label{conv-t4}\\
				&	\hspace{6cm}   
					[ \VN( x_j-x_{k+1}, x_j-x_{k+2} ) , \widetilde\gamma_{N,s}^{(k+2)} ]. 
                  \nonumber
\end{align} 
Now we observe the following: 
\begin{itemize} 
\item As $N \rightarrow \infty$, the term \eqref{conv-lhs} converges to the term 
on the lhs of \eqref{conv-suff2}, due to \eqref{conv-assum}.
\item Also thanks to \eqref{conv-assum}, the term \eqref{conv-t1} converges to 
the first term on the rhs of \eqref{conv-suff2}.
\item The terms \eqref{conv-t2} and \eqref{conv-t3} vanish as $N \rightarrow \infty$.
\end{itemize} 
Hence, it suffices to prove that \eqref{conv-t4} converges to the last term on the rhs of 
\eqref{conv-suff2}, as $N \rightarrow \infty$. Also since the contributions in \eqref{conv-t4} 
proportional to $\frac{k(k-1)}{N^2}$ as well as those proportional to $\frac{k}{N}$ and
to $\frac{k-1}{N}$ vanish as $N \rightarrow \infty$, we only need to prove that, 
for fixed $T$, $k$ and $J^{(k)}$ we have: 
\begin{align}  
             \sup_{s \leq t \leq T} & |\tr J^{(k)} \cU^{(k)}(t-s) \nonumber \\ 
             & \left( \VN( x_j-x_{k+1}, x_j-x_{k+2} ) \widetilde\gamma_{N,s}^{(k+2)}
             - b_0 \delta(x_j - x_{k+1}) \delta(x_j - x_{k+2}) 
             \gamma^{(k+2)}_{\infty,s} \right)|
             \rightarrow 0, \label{conv-suff3}
\end{align}
as $N \rightarrow \infty$. We shall prove \eqref{conv-suff3} following the ideas presented
in \cite{kiscst}. More precisely, let us choose a probability measure $h \in L^1({\R}^d)$, such that
$h \geq 0$ and $\int h = 1$. For an arbitrary $a > 0$, we define 
$h_a(x) = \frac{1}{a^d} h(\frac{x}{a})$. Also let us introduce the notation 
$J^{(k)}_{s-t} := J^{(k)} \cU^{(k)}(t-s)$. Then we have: 
\begin{align} 
       |\tr & J^{(k)}_{s-t}  \left( \VN( x_j-x_{k+1}, x_j-x_{k+2} ) \widetilde\gamma_{N,s}^{(k+2)} - b_0 \delta(x_j - x_{k+1}) \delta(x_j - x_{k+2})  
                                    \gamma^{(k+2)}_{\infty,s} \right)| \nonumber \\
                                 & \leq  |\tr J^{(k)}_{s-t} \left( \VN( x_j-x_{k+1}, x_j-x_{k+2} ) - b_0 \delta(x_j - x_{k+1}) \delta(x_j - x_{k+2})  \right)
                                    \widetilde\gamma_{N,s}^{(k+2)}| \label{conv-exp1}  \\
                                 & + b_0|\tr J^{(k)}_{s-t} \left( \delta(x_j - x_{k+1}) \delta(x_j - x_{k+2})  - h_{a}(x_j - x_{k+1}) h_{a}(x_j - x_{k+2}) \right)  
                                     \widetilde\gamma_{N,s}^{(k+2)}| \label{conv-exp2} \\  
                                 & + b_0|\tr J^{(k)}_{s-t} \; h_{a}(x_j - x_{k+1}) h_{a}(x_j - x_{k+2}) 
                                      \left( \widetilde \gamma_{N,s}^{(k+2)} -  \gamma_{\infty,s}^{(k+2)} \right) |\label{conv-exp3} \\
                                 & + b_0|\tr J^{(k)}_{s-t} \left( h_{a}(x_j - x_{k+1}) h_{a}(x_j - x_{k+2}) - b_0 \delta(x_j - x_{k+1}) \delta(x_j - x_{k+2})  \right)
                                    \gamma_{\infty,s}^{(k+2)}|. \label{conv-exp4}  
\end{align}            
Now we observe that:
\begin{itemize} 
\item The term \eqref{conv-exp1} converges to zero as $N \rightarrow \infty$ by Lemma \ref{lemma-poincare} and by the a priori bounds  \eqref{apriori-1-boundNenergy}.
\item The term \eqref{conv-exp2} converges to zero as $a \rightarrow 0$, uniformly in $N$, by Lemma \ref{lemma-poincare} and by the a priori bounds  \eqref{apriori-1-boundNenergy}.
\item The term \eqref{conv-exp3} converges to zero as $N \rightarrow \infty$, for every fixed $a$  thanks to calculations that are analogous to (6.8) in \cite{kiscst}. 
\item The term \eqref{conv-exp4} converges to zero as $a \rightarrow 0$ by Lemma \ref{lemma-poincare} and by the a priori bounds \eqref{bd-limtrace} on the limiting hierarchy 
(obtained in Proposition \ref{prop-apriori-bound-limhier}). 
\end{itemize} 
Now we first take the limit as $N \rightarrow \infty$ in \eqref{conv-exp1} - \eqref{conv-exp4}, and then let $a\rightarrow 0$. This proves \eqref{conv-suff3}. 
\endprf

The cutoff parametrized by $\kappa>0$ that is introduced in \eqref{defcutof}
can be removed by the same limiting procedure as in \cite{esy}, see also \cite{kiscst}.
We quote the main steps for the convenience of the reader, from \cite{esy,kiscst}.

For the limiting hierarchy $\widetilde\Gamma_{N,t}\rightarrow\Gamma_{\infty,t}$ as $N\rightarrow\infty$, 
it is proven below that for every $\kappa>0$, 
$\widehat\eta(\widehat\gamma_{N,t}^{(k)} , |\phi_t\rangle\langle \phi_t|^{\otimes k})\rightarrow0$
as $N\rightarrow\infty$, for every fixed $k$.
This also implies the convergence 
\eqn\label{eq-hatgamma-conv-1}
	\widetilde\gamma_{N,t}^{(k)} \, \rightarrow \,
	|\phi_t\rangle\langle \phi_t|^{\otimes k}
\eeqn 
in the weak* topology of $\cL_k^1$.

It remains to be proven that also $\gamma_{N,t}^{(k)}\rightarrow 
|\phi_t\rangle\langle \phi_t|^{\otimes k}$. 
To this end, one may assume $\kappa>0$ to be
sufficiently small such that
\eqn 
 	\Big| \, \tr J^{(k)} \, \big( \,  \gamma_{N,t}^{(k)} 
	- \widetilde\gamma_{N,t}^{(k)} \, \big) \, \Big| \, \leq \,
	\| \, J^{(k)} \, \| \, \| \, \Psi_N - \widetilde \Psi_N \, \| \, < \, C \kappa 
	\, \leq \, \frac\e2 \,,
\eeqn
uniformly in $N$. 
This follows from $\| \Psi_N - \widetilde \Psi_N \| < C \kappa $,
uniformly in $N$, which can be easily verified.
On the other hand, for all $N>N_0$ with $N_0$ sufficiently large, we have
\eqn 
 	\Big| \, \tr J^{(k)} \, \big( \, \widetilde\gamma_{N,t}^{(k)} 
	-  |\phi_t\rangle\langle \phi_t|^{\otimes k} \, \big) \, \Big| \, \leq \,  \frac\e2 \,,
\eeqn
due to the convergence of $\widetilde\gamma_{N,t}^{(k)}$ described above.
This implies that for arbitrary $\e>0$,
\eqn 
 	\Big| \, \tr J^{(k)} \, \big( \,  \gamma_{N,t}^{(k)} 
	-  |\phi_t\rangle\langle \phi_t|^{\otimes k} \, \big) \, \Big| \, \leq \, \e \,,
\eeqn
for all $N>N_0$. Thus, for every $t\in[0,T]$ and every fixed $k$, $\gamma_{N,t}^{(k)} \rightarrow 
|\phi_t\rangle\langle \phi_t|^{\otimes k}$ in the weak* topology of $\cL_k^1$.
Because the limiting density is an orthogonal projection, this is equivalent to 
the convergence in trace norm topology. For details, we refer to \cite{esy,kiscst}.

\section{A priori energy bounds on the limiting hierarchy}
\label{sect-boundsGP-1}

In this section we prove some spatial bounds for the limit points
$\{ \gamma^{(k)}_{\infty,t} \}_{k \geq 1}$ that shall be used in order to 
prove uniqueness of the hierarchy.

More precisely, first we state the a-priori bound which follows from 
the estimates \eqref{bd-trace} for $\tilde{\gamma}^{(k)}_{N,t}$.

\begin{proposition} \label{prop-apriori-bound-limhier}
If 
$\Gamma_{\infty,t} = \{ \gamma_{\infty,t}^{(k)} \}_{k \geq 1}$ 
is a limit point of the sequence 
$\tilde{\Gamma}_{N,t} = \{\tilde{\gamma}_{N,t}^{(k)} \}_{k=1}^{N}$
with respect to the product topology $\tau_{prod}$, then there exists
$C>0$ such that 
\eqn \label{bd-limtrace}
\tr(1-\Delta_1) \cdots (1-\Delta_k) \gamma_{\infty,t}^{(k)} \leq C^k,
\eeqn
for all $k \geq 1$. 
\end{proposition}  
\prf 
The proof follows from the fact that the a-priori estimates 
\eqref{bd-trace} for $\tilde{\gamma}^{(k)}_{N,t}$ hold 
uniformly in $N$.   
\endprf

As in \cite{kiscst},  we prove uniqueness of the infinite hierarchy 
following the approach introduced by Klainerman and Machedon \cite{klma}.
In order to apply the approach of \cite{klma}  we establish another a-priori 
bound on the limiting density. Such a bound is formulated in Theorem \ref{KMbound}
below. In what follows  $S^{(k,\alpha)}$ denotes 
$$ S^{(k,\alpha)} = \prod_{j=1}^{k}
(1-\Delta_{x_j})^{\frac{\alpha}{2}} (1 - \Delta_{x'_j})^{\frac{\alpha}{2}}.$$

\begin{theorem} \label{KMbound} 
Suppose that $d \in \{1,2\}$. If   
$\Gamma_{\infty,t} = \{ \gamma_{\infty,t}^{(k)} \}_{k \geq 1}$ 
is a limit point of the sequence 
$\tilde{\Gamma}_{N,t} = \{\tilde{\gamma}_{N,t}^{(k)} \}_{k=1}^{N}$
with respect to the product topology $\tau_{prod}$, then, 
for every $\alpha < 1$ if $d=2$, and every $\alpha\leq1$ if $d=1$, there exists $C>0$ such that 
\eqn \label{bd-KM} 
\|S^{(k,\alpha)} B_{j;k+1,k+2} \gamma^{(k+2)}_{\infty,t} \|_{L^2({\R}^{dk} \times {\R}^{dk})} 
\leq C^{k+2},
\eeqn
for all $k \geq 1$ and all $t \in [0,T]$. 
\end{theorem} 

\prf 
We modify the proof of an analogous result presented in Theorem 5.2 of \cite{kiscst}. 
We note that for the argument employed here, the fact is used that $\gamma^{(\ell)}_{\infty,t}$
is positive, and thus, especially, hermitean. 
We note that Theorem \ref{thm-spacetime-bd-1} below states a similar result,
but for a different quantity than  
$\gamma^{(\ell)}_{\infty,t}$ which may be neither positive nor hermitean. 
Thus, the proof of  Theorem \ref{thm-spacetime-bd-1}
is based on a different approach that necessitates a {\em lower} bound on $\alpha$,
instead of an upper bound as required here.

By \eqref{bd-limtrace} it suffices to prove 
\eqn \label{bd-KMtrace} 
	\|S^{(k,\alpha)} B_{j;k+1,k+2} \gamma^{(k+2)}_{\infty,t} \|_{L^2({\R}^{dk} \times {\R}^{dk})} 
	\, \leq \, 
	\tr(1-\Delta_1)...(1-\Delta_{k+2}) \gamma_{\infty,t}^{(k+2)}.
\eeqn

We will consider the case $k=1, j=1$ (the argument for $k \geq 2$ 
can be carried out in a similar way). We start by calculating the Fourier transform
of $B_{1;2,3} \gamma_{\infty,t}^{(3)}$. It suffices to do that for  $B^{+}_{1;2,3} \gamma_{\infty,t}^{(3)}$ (the calculations 
for  $B^{-}_{1;2,3} \gamma_{\infty,t}^{(3)}$ can be carried out in an analogous way):

\eqn
             \lefteqn{
             \widehat{ B^{+}_{1;2,3} \gamma_{\infty,t}^{(3)} }(p;p') 
             }
              \nonumber\\
             & = & \int dx_1 \; dx'_1 e^{-i x_1 \cdot p} e^{i x'_1 \cdot p'}     
                  \int dx_2 \; dx'_2 \; dx_3 \; dx'_3 
              \nonumber\\ 
                && \delta(x_1 - x_2) \delta(x_1 - x'_2) \delta(x_1 - x_3) \delta(x_1 - x'_3) 
                  \gamma^{(3)}_{\infty,t}(x_1,x_2,x_3;x'_1,x'_2,x'_3) 
              \nonumber\\ 
              & = & \int dq \; d\kappa \; dr\; ds 
                  \int dx_1 \; dx'_1 \; dx_2 \; dx'_2 \; dx_3 \; dx'_3 
              \nonumber\\ 
                && e^{-i x_1 \cdot p} e^{i x'_1 \cdot p'} 
                  e^{iq(x_1-x_2)} e^{-i\kappa(x_1-x'_2)} 
                  e^{ir(x_1-x_3)} e^{-is(x_1-x'_3)} 
                  \gamma^{(3)}_{\infty,t}(x_1,x_2,x_3;x'_1,x'_2,x'_3) 
              \nonumber\\ 
              & = & \int dq \; d\kappa \; dr\; ds 
                  \int dx_1 \; dx'_1 \; dx_2 \; dx'_2 \; dx_3 \; dx'_3 
              \nonumber\\
                && e^{-ix_1 \cdot (p - q + \kappa - r + s)} 
                  e^{-ix_2 \cdot q}  e^{-ix_3 \cdot r}  
                  e^{ix'_1 \cdot p'} e^{ix'_2 \cdot \kappa }e^{ix'_3 \cdot s} 
                  \gamma^{(3)}_{\infty,t}(x_1,x_2,x_3;x'_1,x'_2,x'_3) 
              \nonumber\\
              & = & \int dq \; d\kappa \; dr\; ds \;                
                  \widehat{ \gamma^{(3)}_{\infty,t} }(p - q + \kappa -r + s,q,r;p',\kappa,s).    
\eeqn
Hence 
\eqn
  \lefteqn{
  \widehat{ S^{(1,\alpha)} B^{+}_{1;2,3} \gamma^{(3)}_{\infty,t} }(p;p') 
  }
 \\
  &&= \, \bra p \ket^\alpha \bra p' \ket^\alpha  
  \int dq \; d\kappa \; dr\; ds\;                
  \widehat{ \gamma^{(3)}_{\infty,t} }(p - q + \kappa - r + s,q,r;p',\kappa,s) \, ,
  \nonumber
\eeqn 
which in turn implies 
\eqn
             \lefteqn{
             \| S^{(1,\alpha)} B^{+}_{1;2,3} \gamma^{(3)}_{\infty,t} (p;p') \|
             _{L^2({\R}^{d} \times {\R}^{d})}^2
             }
             \nonumber\\
            &=& \int dp\; dp' \; dq_1 \; dq_2 \; d\kappa_1 \; 
               d\kappa_2 \; dr_1 \; dr_2 \; ds_1 \; ds_2  
               \bra p \ket^{2\alpha} \bra p' \ket^{2\alpha}  \nonumber \\ 
            & & \quad \quad \quad 
               \widehat{ \gamma^{(3)}_{\infty,t} }(p - q_1 + \kappa_1 - r_1 + s_1,q_1,r_1;p'_1,\kappa_1,s_1) 
               \nonumber \\ 
            & & \quad \quad \quad \quad \quad
               \widehat{ \gamma^{(3)}_{\infty,t} }(p-q_2+\kappa_2-r_2+s_2,q_2,r_2;p'_2,\kappa_2,s_2).
               \label{norm1al-afterF}
\eeqn
Substituting 
\eqn\label{eq-gamma-eigenvexp-1}
 \widehat{ \gamma^{(3)}_{\infty,t}}(p_1,p_2,p_3;p'_1,p'_2,p'_3) = \sum_j \lambda_j 
  \psi_j(p_1,p_2,p_3) \bar{\psi}_{j}(p'_1,p'_2,p'_3)
\eeqn 
into \eqref{norm1al-afterF} and keeping in mind that $\lambda_j \geq 0$ for all $j$ 
and $\sum_j \lambda_j \leq 1$ thanks to $\gamma^{(k+2)}$ being a non-negative 
trace-class operator with trace at most one,  
we obtain: 
\eqn
            \lefteqn{
             \| S^{(1,\alpha)} B^{+}_{1;2,3} \gamma^{(3)}_{\infty,t} (p;p') \|
             _{L^2({\R}^{d} \times {\R}^{d})}^2  
             }
             \nonumber\\
           & = & \sum_{i,j} \lambda_i \lambda_j s
               \int dp\; dp' \; dq_1 \; dq_2 \; d\kappa_1 \; 
               d\kappa_2 \; dr_1 \; dr_2 \; ds_1 \; ds_2  
               \bra p \ket^{2\alpha} \bra p'\ket^{2\alpha} \nonumber \\ 
             &&\quad\quad\quad\quad 
               \psi_j(p-q_1+\kappa_1-r_1+s_1,q_1,r_1) 
               \bar{\psi}_j(p',\kappa_1,s_1) \nonumber \\ 
             &&\quad\quad\quad\quad\quad\quad
               \psi_j(p-q_2+\kappa_2-r_2+s_2,q_2,r_2) 
               \bar{\psi}_j(p',\kappa_2,s_2). \label{usepsi}  
\eeqn
We observe that for $l=1,2$ we have 
$$ 
  \bra p \ket^\alpha  \leq C 
  \left[ 
  \bra p-q_l+\kappa_l -r_l + s_l \ket^\alpha   
  + \bra q_l \ket^\alpha  
  + \bra r_l \ket^\alpha 
  + \bra \kappa_l \ket^\alpha 
  + \bra s_l \ket^\alpha 
  \right]
$$
which implies that 
\eqn\label{prod}
              \bra p\ket^{2\alpha}  
              & \leq & C  
              \left[ 
              \bra p - q_1+\kappa_1 - r_1 + s_1 \ket^{ \alpha }  
              + \bra q_1 \ket^\alpha 
              + \bra r_1 \ket^\alpha 
              + \bra \kappa_1 \ket^\alpha 
              + \bra s_1 \ket^\alpha 
              \right] \\
              && \quad\quad\quad\times 
              \left[ 
              \bra p - q_2+\kappa_2 - r_2 + s_2 \ket^\alpha 
              + \bra q_2 \ket^\alpha  
              + \bra r_2 \ket^\alpha 
              + \bra \kappa_2 \ket^\alpha 
              + \bra s_2 \ket^\alpha 
              \right].  
              \nonumber 
\eeqn
Substituting \eqref{prod} into \eqref{usepsi}, we obtain $16$ terms. We will
illustrate how to control one of them, the remaining cases are similar. 
Using a weighted Schwarz inequality, we find 
\eqnn 
              \lefteqn{
               \int dp\; dp' \; dq_1 \; dq_2 \; d\kappa_1 \; 
                d\kappa_2 \; dr_1 \; dr_2 \; ds_1 \; ds_2  
                }
                \\
              &&
                \bra p' \ket^{2\alpha}  
                \bra p-q_1+\kappa_1 - r_1 + s_1 \ket^\alpha   
                \bra p-q_2+\kappa_2 - r_2 + s_2 \ket^\alpha  
                \nonumber \\ 
              && \quad\quad 
               \psi_j(p-q_1+\kappa_1-r_1+s_1,q_1,r_1) 
               \bar{\psi}_j(p',\kappa_1,s_1) \\ 
              && \quad\quad\quad\quad
               \psi_j(p-q_2+\kappa_2-r_2+s_2,q_2,r_2) 
               \bar{\psi}_j(p',\kappa_2,s_2) \\ 
         \leq & I + II, 
\eeqnn
where
\begin{align*} 
              I =  \int & dp\; dp' \; dq_1 \; dq_2 \; d\kappa_1 \; 
                     d\kappa_2 \; dr_1 \; dr_2 \; ds_1 \; ds_2  
                     \\
                   & \quad\quad
                     \frac{ \bra {p'} \ket^{2\alpha} 
                            \bra p-q_1+\kappa_1-r_1+s_1\ket^2 \bra q_1 \ket^2 \bra r_1 \ket^2 
                            \bra \kappa_2 \ket^2 \bra s_2 \ket^2  }     
                          { \bra p-q_2+\kappa_2-r_2+s_2)^2 \ket^{2-2\alpha} 
                            \bra q_2 \ket^2 \bra r_2 \ket^2 \bra \kappa_1 \ket^2 \bra s_1 \ket^2 } \\
                   & \quad\quad\quad\quad
                     \left|\psi_j(p-q_1+\kappa_1-r_1+s_1,q_1,r_1)\right|^2 
                     \left|{\psi}_j(p',\kappa_2,s_2)\right|^2,
\end{align*}
and 
\begin{align*} 
              II =  \int & dp\; dp' \; dq_1 \; dq_2 \; d\kappa_1 \; 
                     d\kappa_2 \; dr_1 \; dr_2 \; ds_1 \; ds_2  
                     \\
                   & \quad\quad 
                     \frac{ \bra p' \ket^{2\alpha}  
                            \bra p-q_2+\kappa_2-r_2+s_2 \ket^2 \bra q_2 \ket^2 \bra r_2 \ket^2 
                            \bra \kappa_1 \ket^2 \bra s_1 \ket^2 }     
                          { \bra p-q_1+\kappa_1-r_1+s_1 \ket^{2-2\alpha} 
                            \bra q_1 \ket^2 \bra r_1 \ket^2 \bra \kappa_2 \ket^2 \bra s_2 \ket^2 } \\
                   & \quad\quad\quad\quad
                     \left|\psi_j(p-q_2+\kappa_2-r_2+s_2,q_2,r_2)\right|^2 
                     \left|{\psi}_j(p',\kappa_1,s_1)\right|^2.
\end{align*}
Below we illustrate how to estimate $I$. The expression $II$ can be estimated in a similar 
manner. We will use the bound  
\eqn \label{crucialint} 
                       \int_{{\R}^d} 
                       \frac{dy}{\bra P-y \ket^{2-2\alpha} \bra y \ket^2 } 
                       \leq \frac{C}{\bra P \ket^{2-2\alpha}},  
\eeqn
which is valid for $d = 1$ if $\alpha\leq1$, and for $d=2$ if $\alpha<1$;
it is easily obtained by rescaling $y\rightarrow\bra P\ket y$. 
To estimate $I$, we integrate over $q_2$, using \eqref{crucialint}, 
followed by integrating over $r_2$, using \eqref{crucialint} again, to 
obtain:   
\begin{align*} 
              I \leq \int & dp\; dp' \; dq_1 \; d\kappa_1 \; 
                     d\kappa_2 \; dr_1 \; ds_1 \; ds_2  
                     \\
                   & \quad\quad 
                     \frac{ \bra p' \ket^{2\alpha} 
                            \bra p-q_1+\kappa_1-r_1+s_1 \ket^2 \bra q_1 \ket^2 \bra r_1 \ket^2  
                            \bra \kappa_2 \ket^2 \bra s_2 \ket^2 }     
                          { \bra p+\kappa_2+s_2 \ket^{2-2\alpha} 
                            \bra \kappa_1 \ket^2 \bra s_1 \ket^2 } \\
                   & \quad\quad\quad\quad
                    \left|\psi_j(p-q_1+\kappa_1-r_1+s_1,q_1,r_1)\right|^2 
                     \left|{\psi}_j(p',\kappa_2,s_2)\right|^2 \,.
\end{align*}
The change of variable 
$\tilde{p} = p - q_1 +\kappa_1 - r_1 + s_1$ gives
\eqn
              I &\leq & \int   d\tilde{p}\; dp' \; dq_1 \; d\kappa_1 \; 
                     d\kappa_2 \; dr_1 \; ds_1 \; ds_2 \nonumber \\ 
                   &&\quad\quad 
                     \frac{ \bra {\tilde{p}} \ket^2 \bra q_1 \ket^2 \bra r_1 \ket^2  
                            \bra {p'} \ket^2 \bra \kappa_2 \ket^2 \bra s_2 \ket^2 }     
                          { \bra \tilde{p} + q_1 - \kappa_1 + r_1 - s_1 
                                  + \kappa_2 + s_2 \ket^{2-2\alpha} 
                            \bra \kappa_1 \ket^2 \bra s_1 \ket^2 } \nonumber \\
                   &&\quad\quad\quad\quad\quad\quad\quad\quad\quad\quad\quad\quad\quad 
					\left|\psi_j(\tilde{p},q_1,r_1)\right|^2 
					\left|{\psi}_j(p',\kappa_2,s_2)\right|^2 \nonumber \\
              & \leq & C_{\alpha} 
                     \int d\tilde{p} \; dq_1 \; dr_1 \,
                     \bra {\tilde{p}} \ket^2 \bra q_1 \ket^2 \bra r_1 \ket^2   
                     \left|\psi_j(\tilde{p},q_1,r_1)\right|^2 \nonumber \\ 
                   &&\quad\quad\quad\quad
                    \int dp' \; d\kappa_2 \; ds_2 \,
                     \bra {p'} \ket^2 \bra \kappa_2 \ket^2 \bra s_2 \ket^2   
                     \left|\psi_j(p',\kappa_2,s_2)\right|^2 \,. \label{useCal}                    
\eeqn
To obtain \eqref{useCal} we have used  that,  
as a consequence of \eqref{crucialint},
\eqn \label{Cal}  
  C_{\alpha} 
  \, = \, \sup_{P \in {\R}^{d}} \int 
  \frac{dy \, dz}{ \bra P-y-z \ket^{2-2\alpha} \bra y \ket^2 \bra z \ket^2  } 
  \, < \, \infty \, ,
\eeqn
for all $\alpha\leq1$ if $d=1$, and all  $\alpha < 1$ if $d=2$.  

The other $15$ contributions to \eqref{usepsi} can be obtained in a similar way. 
Therefore, using the above analysis and \eqref{usepsi}, we conclude that
\eqn
              \lefteqn{
              \| S^{(1,\alpha)} B^{+}_{1;2,3} \gamma^{(3)}_{\infty,t} (p;p') \|
              _{L^2({\R}^{d} \times {\R}^{d})}^2 
              }
              \nonumber\\ 
              &\leq & C \sum_{i,j} \lambda_i \lambda_j 
                     \int d\tilde{p} \; dq_1 \; dr_1 
                     \bra {\tilde{p}} \ket^2 \bra q_1 \ket^2 \bra r_1 \ket^2  
                     \left|\psi_j(\tilde{p},q_1,r_1)\right|^2 
              \nonumber\\ 
              &     & \quad\quad\quad\quad
                     \int dp' \; d\kappa_2 \; ds_2 
                     \bra {p'} \ket^2 \bra \kappa_2 \ket^2 \bra s_2 \ket^2  
                     \left|\psi_j(p',\kappa_2,s_2)\right|^2 
              \nonumber\\
              &\leq & \left[ \, \int d\tilde{p} \; dq_1 \; dr_1 
                     \bra {\tilde{p}} \ket^2 \bra q_1 \ket^2 \bra r_1 \ket^2   
                     \left| \widehat{\gamma^{(3)}_{\infty,t}} (\tilde{p},q_1,r_1; \tilde{p}, q_1, r_1)\right|^2 \, \right]^2 
              \nonumber\\         
              &=    & C \left[ \, \tr(1-\Delta_1)(1-\Delta_2)(1-\Delta_3) \, \gamma^{(3)}_{\infty,t} \, \right]^2,
\eeqn
which gives \eqref{bd-KMtrace} in the case $k=1$, $j=1$.  
\endprf

In addition to the above results, 
we derive a third type of spatial bounds, which is more restrictive in terms 
of the condition on $\alpha$ (it requires $\alpha > \frac{d}{2}$). Note that for  
$d =1 $ we can afford this range of  $\alpha$. In particular, we shall use 
this new bound iteratively in the proof of uniqueness of the limiting hierarchy when $d=1$.  
The proof of the bound is inspired by 
the proof of a space-time bound for the freely evolving limiting hierarchy  
given in Theorem 1.3 of \cite{klma}. However, the bound that we derive here 
is obtained for any $\gamma^{(k)}_{\infty,t}$. 

\begin{theorem} \label{highregbound} 
Suppose that $d  \geq 1$. If   
$\Gamma_{\infty,t} = \{ \gamma_{\infty,t}^{(k)} \}_{k \geq 1}$ 
is a limit point of the sequence 
$\tilde{\Gamma}_{N,t} = \{\tilde{\gamma}_{N,t}^{(k)} \}_{k=1}^{N}$
with respect to the product topology $\tau_{prod}$, then, 
for every $\alpha>\frac d2$ there exists a constant $C=C(\alpha)$ such that
the estimate
\eqn\label{eq-aprioribd-quintic-1}
	\lefteqn{
	\Big\| \, S^{(k,\alpha)}B_{j;k+1,k+2}\gamma^{(k+2)}_{\infty,t} \, 
	\Big\|_{ L^2(\R^{dk}\times\R^{dk})}
	}
	\nonumber\\
	&&\quad\quad\quad\quad
	\, \leq \, C \, 
	\Big\| \,  S^{(k+2,\alpha)} \gamma^{(k+2)}_{\infty,t} \, \Big\|_{L^2(\R^{d(k+2)}\times\R^{d(k+2)})}  
\eeqn 
holds. 
\end{theorem}
 
\prf
Let $(\uu_k,\uu_k')$, $\uq:=(q_1,q_2)$, and $\uq':=(q_1',q_2')$
denote the Fourier conjugate variables corresponding to 
$(\ux_k,\ux_k')$, $(x_{k+1},x_{k+2})$, and $(x_{k+1}',x_{k+2}')$, respectively.

Without any loss of generality, we may assume that $j=1$ in $B_{j;k+1,k+2}$. Then, 
we have 
\eqn
	\lefteqn{
	\Big\| \, S^{(k,\alpha)}B_{1;k+1,k+2}\gamma^{(k+2)}_{\infty,t} \, 
	\Big\|_{L^2( \R^{ dk}\times\R^{ dk})}^2
	}
	\nonumber\\
	& = &
	\int d\uu_k \, d\uu_k' \prod_{j=1}^k\bra u_j\ket^{2\alpha} \bra u_j'\ket^{2\alpha}
	\\
	&&\quad\quad\quad
	\Big( \, \int d\uq \, d\uq' \,  
	\widehat\gamma^{(k+2)}_{\infty,t}(u_1+q_1+q_2-q_1'-q_2',u_2,\dots,u_k,\uq;\uu_k',\uq')\, \Big)^2 \,.
	\nonumber
\eeqn
where now, the Fourier transform in only performed in the spatial coordinates.
Applying the Schwarz inequality, we find the upper bound
\eqn
	&\leq& \int d\uu_k \, d\uu_k' \, I_\alpha'(\tau,\uu_k,\uu_k')
	\int d\uq \, d\uq' \,  
 	\nonumber\\
	&&
	\bra u_1+q_1+q_2-q_1'-q_2'\ket^{2\alpha} \bra q_1\ket^\alpha\bra q_2\ket^{2\alpha}
	\bra q_1'\ket^\alpha\bra q_2'\ket^{2\alpha}
	\prod_{j=2}^k\bra u_j\ket^{2\alpha} \prod_{j'=1}^k\bra u_{j'}'\ket^{2\alpha}
	\nonumber\\
	&& \quad \quad \quad \quad \quad \quad  
	\Big| \, \widehat\gamma^{(k+2)}_{\infty,t}(u_1+q_1+q_2-q_1'-q_2',u_2,\dots,u_k,\uq;\uu_k',\uq') \, \Big|^2
\eeqn
where
\eqn\label{eq-Ialpha-def-2}
	\lefteqn{
	I_\alpha'(\uu_k,\uu_k') 
	}
	\\
	&&
	\, := \,
	\int d\uq d\uq' \, 
	\frac{ \bra u_1\ket^{2\alpha}}
	{\bra u_1+q_1+q_2-q_1'-q_2'\ket^{2\alpha} \bra q_1\ket^{2\alpha} \bra q_2\ket^{2\alpha}
	\bra q_1'\ket^{2\alpha} \bra q_2'\ket^{2\alpha}} \,.
	\nonumber
\eeqn
Using
\begin{equation}\label{eq-u1-aux-bd-1} 
	 \bra u_1\ket^{2\alpha}  
	 \, \leq \, 
	 C\Big[ \, \bra u_1+q_1+q_2-q_1'-q_2'\ket^{2\alpha} 
	 + \bra q_1\ket^{2\alpha}
	 + \bra q_2\ket^{2\alpha}
	 + \bra q_1'\ket^{2\alpha} + \bra q_2'\ket^{2\alpha} \, \Big]
	 \,, 
\end{equation} 
and shifting some of the momentum variables, one immediately 
obtains that
\eqn\label{eq-IalphaA-bd-1}
	I_\alpha'(\uu_k,\uu_k') \, < \, C \, 
	\int d\uq d\uq' \, 
	\frac{  1 }
	{\bra q_1\ket^{2\alpha} \bra q_2\ket^{2\alpha}
	\bra q_1'\ket^{2\alpha} \bra q_2'\ket^{2\alpha}} \,,
\eeqn
which is finite for all 
\eqn
	\alpha \, > \, \frac d2 \,.
\eeqn
This proves the claim.
\endprf

\section{Bounds on the freely evolving infinite hierarchy}
\label{sect-boundsfreeGP-1}

In this section, we prove bounds on the infinite hierarchy for $b_0=0$,
i.e., in the absence of particle interactions; see (\ref{eq-def-b0-1}) for the definition of $b_0$. 
These will be used for the recursive estimation of terms appearing in
the Duhamel expansions studied in Section \ref{sect-Duhamel-1}.
Our approach is similar to the one of Klainerman and Machedon in \cite{klma}.
In dimension $d=2$, we prove spacetime bounds in complete analogy to
\cite{klma,kiscst} which are global in time \footnote{In dimension $d=1$, the argument used for $d=2$ would produce a divergent
bound; accordingly, when $d=1$ we shall use the a priori bounds obtained in 
Theorem \ref{highregbound}.}.

From here on and for the rest of this paper, we will write
\eqn
	\gamma^{(r)}(t,\ux_k;\ux_k') \, \equiv \, \gamma^{(r)}_{\infty,t}(t,\ux_k;\ux_k')
\eeqn
which is notationally more convenient for the discussion of spacetime norms.

\begin{theorem}
\label{thm-spacetime-bd-1}
Assume that $d=2$ and $\frac56<\alpha<1$. 
Let $\gamma^{(k+2)}$ denote the solution of 
\eqn
	i\partial_t\gamma^{(k+2)}(t,\ux_{k+2};\ux_{k+2}')
	\, + \, (\Delta_{\ux_{k+2}}-\Delta_{\ux_{k+2}'})
	\gamma^{(k+2)}(t,\ux_{k+2};\ux_{k+2}') \, = \, 0
\eeqn
with initial condition
\eqn
	\gamma^{(k+2)}(0,\,\cdot\,) \, = \, \gamma_0^{(k+2)}\in\cH^\alpha \,.
\eeqn 
Then, there exists a constant $C=C(\alpha)$ such that 
\eqn
	\lefteqn{
	\Big\| \, S^{(k,\alpha)}B_{j;k+1,k+2}\gamma^{(k+2)} \, 
	\Big\|_{L^2_{t,\ux_{k},\ux_{k}'}(\R\times\R^{2k}\times\R^{2k})}
	}
	\nonumber\\
	&&\quad\quad\quad\quad
	\, \leq \, C \, 
	\Big\| \,  S^{(k+2,\alpha)} \gamma_0^{(k+2)} \, 
	\Big\|_{L^2_{\ux_{k+2},\ux_{k+2}'}(\R^{2(k+2)}\times\R^{2(k+2)})}  
\eeqn 
holds.
\end{theorem}

\prf
We give a proof  using the arguments of \cite{klma,kiscst}. 
We note that the arguments presented in the proof of Theorem \ref{KMbound} 
cannot be straightforwardly employed here because here,
$B_{j;k+1,k+2}\gamma^{(k+2)}$ are not hermitean so that
(\ref{eq-gamma-eigenvexp-1}) is not available.

Let $(\tau,\uu_k,\uu_k')$, $\uq:=(q_1,q_2)$, and $\uq':=(q_1',q_2')$
denote the Fourier conjugate variables corresponding to 
$(t,\ux_k,\ux_k')$, $(x_{k+1},x_{k+2})$, and $(x_{k+1}',x_{k+2}')$, respectively.

Without any loss of generality, we may assume that $j=1$ in $B_{j;k+1,k+2}$. Then, 
abbreviating
\eqn
	\delta(\cdots) \, := \, \delta( \, \tau + (u_1+q_1+q_2-q_1'-q_2')^2  
	+ \sum_{j=2}^k u_j^2
	+ |\uq|^2 
	- |\uu_k'|^2 - |\uq'|^2   \, )
\eeqn
we find
\eqn
	\lefteqn{
	\Big\| \, S^{(k,\alpha)}B_{1;k+1,k+2}\gamma^{(k+2)} \, 
	\Big\|_{L^2_{t,\ux_{k},\ux_{k}'}(\R\times\R^{2(k+2)}\times\R^{2(k+2)})}^2
	}
	\nonumber\\
	& = &
	\int_{\R}d\tau \int d\uu_k d\uu_k' \prod_{j=1}^k\bra u_j\ket^{2\alpha} \bra u_j'\ket^{2\alpha}
	\\
	&&\quad 
	\Big( \, \int d\uq d\uq' \, 
	\delta(\cdots)
	\widehat\gamma^{(k+2)}(\tau,u_1+q_1+q_2-q_1'-q_2',u_2,\dots,u_k,\uq;\uu_k',\uq')\, \Big)^2 \,.
	\nonumber
\eeqn
Using the Schwarz estimate, this is bounded by
\eqn
	&\leq&\int_{\R}d\tau \int d\uu_k d\uu_k' \, I_\alpha(\tau,\uu_k,\uu_k')
	\int d\uq d\uq' \,  
	\delta(\cdots)
	\nonumber\\
	&&
	\bra u_1+q_1+q_2-q_1'-q_2' \ket^{2\alpha} \bra q_1\ket^{2\alpha} \bra q_2\ket^{2\alpha}
	\bra q_1'\ket^{2\alpha} \bra q_2' \ket^{2\alpha}
	\prod_{j=2}^k\bra u_j\ket^{2\alpha} \prod_{j'=1}^k\bra u_{j'}'\ket^{2\alpha}
	\nonumber\\
	&&\quad\quad\quad\quad
	\Big| \, \widehat\gamma^{(k+2)}(\tau,u_1+q_1+q_2-q_1'-q_2',u_2,\dots,u_k,\uq;\uu_k',\uq') \, \Big|^2
\eeqn
where
\eqn\label{eq-Ialpha-def-1}
	\lefteqn{
	I_\alpha(\tau,\uu_k,\uu_k') 
	}
	\\
	&&
	\, := \,
	\int d\uq \, d\uq' \, 
	\frac{
	\delta(\cdots) \, \bra u_1\ket^{2\alpha}}
	{\bra u_1+q_1+q_2-q_1'-q_2'\ket^{2\alpha} \bra q_1\ket^{2\alpha} \bra q_2\ket^{2\alpha}
	\bra q_1'\ket^{2\alpha} \bra q_2'\ket^{2\alpha}} \,.
	\nonumber
\eeqn
Similarly as in \cite{klma,kiscst}, we observe that
\begin{equation}\label{eq-u1-aux-bd-2}
	 \bra u_1\ket^{2\alpha} 
	 \, \leq \, 
	 C\Big[ \, \bra u_1+q_1+q_2-q_1'-q_2'\ket^{2\alpha} 
	 + \bra q_1\ket^{2\alpha}
	 + \bra q_2\ket^{2\alpha}
	 + \bra q_1'\ket^{2\alpha} + \bra q_2'\ket^{2\alpha} \, \Big]
	 \,,
\end{equation} 
so that
\eqn
	I_\alpha(\tau,\uu_k,\uu_k') \, \leq \, \sum_{\ell=1}^5 J_\ell
\eeqn
where $J_\ell$ is obtained from bounding the numerator of
(\ref{eq-Ialpha-def-1}) using  (\ref{eq-u1-aux-bd-2}), 
and from canceling the $\ell$-th term on the rhs 
of (\ref{eq-u1-aux-bd-2}) with the corresponding term
in the denominator of (\ref{eq-Ialpha-def-1}).
Thus, for instance,
\eqn
	J_1 \, < \, \int d\uq \, d\uq' \, 
	\frac{
	\delta(\cdots) }
	{\bra q_1\ket^{2\alpha} \bra q_2\ket^{2\alpha}
	\bra q_1'\ket^{2\alpha} \bra q_2'\ket^{2\alpha}} \,,
\eeqn
and each of the terms $J_\ell$ with $\ell=2,\dots,5$ can be brought into
a similar form by appropriately translating one of the momenta $q_i$, $q_j'$.

Further following \cite{klma,kiscst}, we observe that the argument of the delta distribution
equals
\eqn
	\tau + (u_1+q_1+q_2-q_1')^2  
	+ \sum_{j=2}^k u_j^2
	+ |\uq|^2 
	- |\uu_k'|^2 - (q_1')^2 
	- 2(u_1+q_1+q_2-q_1')\cdot q_2' \,,
	\nonumber 
\eeqn
and we integrate out the delta distribution using the component of $q_2'$ parallel
to $(u_1+q_1+q_2-q_1')$. 
This leads to the bound
\eqn 
	J_1 
	& < &  
	C_\alpha C \int d\uq d q_1' \, 
	\frac{ 1}
	{| u_1+q_1+q_2-q_1'|
	\bra q_1\ket^{2\alpha} \bra q_2\ket^{2\alpha}
	\bra q_1'\ket^{2\alpha} } 
\eeqn
where
\eqn
	C_\alpha \, := \, \int_{\R} \frac{d\zeta}{\bra \zeta \ket^{2\alpha}} \,.
\eeqn 
Clearly, $C_\alpha$ is finite for any $\alpha>\frac12$.

To bound $J_1$, we pick a spherically symmetric  function 
$h\geq0$ with rapid decay away from the unit ball in $\R^2$, such that
$h^\vee(x)\geq0$ decays rapidly outside of the unit ball in $\R^2$, and
\eqn\label{eq-convuppbd-1}
	\frac{1}{\bra q \ket^{2\alpha}} \, < \, \Big( \, h*\frac{1}{| \, \cdot \, |^{2\alpha}} \, \Big)(q) \,.
\eeqn 
(for example, $h(u)= c_1 e^{-c_2 u^2}$, for suitable constants $c_1,c_2$);
since $\alpha<1$, the right hand side is in $L^\infty(\R^2)$. Then, 
\eqn\label{eq-J1-convFTprod-bd-1}
	J_1 
	& < &  
	C_\alpha C \Bra \,
	\Big(\frac{1}{| \, \cdot \, |}*(h*\frac{1}{| \, \cdot \, |^{2\alpha}}) \, \Big) *
	(h*\frac{1}{| \, \cdot \, |^{2\alpha}}) \, , \, (h*\frac{1}{| \, \cdot \, |^{2\alpha}}) \,
	\Ket_{L^2(\R^2)}
	\nonumber\\
	&=& C_\alpha C \int dx \, \Big( \, \frac{1}{| \, \cdot \, |} \, \Big)^\vee(x)
	\, \Big( \, (h*\frac{1}{| \, \cdot \, |^{2\alpha}})^\vee(x) \, \Big)^3
	\nonumber\\
	&=& C_\alpha C' \int dx \,  \frac{1}{| \, x \, |}  (h^\vee(x))^3
	\, \Big( \, \frac{1}{| \, x \, |^{2-2\alpha}} \, \Big)^3 \,.
\eeqn
The integral on the last line is finite if the singularity at $x=0$ is integrable.
In dimension $d=2$, this is the case if
\eqn
	\alpha \, > \, \frac56 \,. 
\eeqn
Finiteness of the integral for the region $|x|\gg1$ is obtained from the decay
of $h^\vee$.  
We remark that if $0<1-\alpha\ll1$, the upper bound (\ref{eq-convuppbd-1}) may
overestimate the left hand side by as much as a factor $\frac{1}{1-\alpha}\gg1$ pointwise in $q$,
for small $|q|$, due to the 
singularity of $\frac{1}{| \, \cdot \,|^{2(1-\alpha)}}$ at zero. 
But the integral in (\ref{eq-J1-convFTprod-bd-1}) is uniformly bounded in the limit $\alpha\nearrow1$,
implying that the argument is robust. 
The terms $J_2,\dots,J_5$ can be bounded in a similar manner.
For more details, we refer to \cite{klma,kiscst}.
This proves the statement of the theorem. 
\endprf

\newpage

\section{Uniqueness of solutions of the infinite hierarchy}
\label{sect-uniq-1}

Collecting our results derived in the previous sections, we now prove
the uniqueness of solutions of the infinite hierarchy.

We recall the notation 
$\Delta_{\pm}^{(k)} = \Delta_{\ux_k} - \Delta_{\ux'_k}$ 
and 
$\Delta_{\pm, x_j} = \Delta_{x_j} - \Delta_{x_j'}$.

Let us fix a positive integer $r$. Using Duhamel's formula we can express 
$\gamma^{(r)}$ in terms of the iterates 
$\gamma^{(r+2)}, \gamma^{(r+4)}, ... , \gamma^{(r+2n)}$ as follows: 

\begin{align} 
\gamma^{(r)}(t_r, \cdot) 
& = \int_{0}^{t_r}  
        e^{ i(t_r - t_{r+2})\Delta_{\pm}^{(r)} } 
      B_{r+2}(\gamma^{(r+2)}(t_{r+2})) \; dt_{r+2} \nonumber \\
& = \int_{0}^{t_r} \int_{0}^{t_{r+2}}
      e^{ i(t_r - t_{r+2})\Delta_{\pm}^{(r)} } 
      B_{r+2} e^{ i(t_{r+2} - t_{r+4})\Delta_{\pm}^{(r+2)} } 
      B_{r+4}(\gamma^{(r+4)}(t_{r+4})) \; dt_{r+2} \; dt_{r+4} \nonumber \\
& = ... \nonumber \\
& = \int_{0}^{t_r} ... \int_{0}^{t_{r+2n-2}}
      J^r(\ut_{r+2n}) \; dt_{r+2} ... dt_{r+2n}, \label{iterates}
\end{align}     
where 
\begin{align*} 
& \ut_{r+2n} = (t_r, t_{r+2},...,t_{r+2n}), \\
& J^r(\ut_{r+2n}) = e^{ i(t_r - t_{r+2})\Delta_{\pm}^{(r)} } 
                    B_{r+2} ... e^{ i(t_{r+2(n-1)} - t_{r+2n})\Delta_{\pm}^{(r+2(n-1))} } 
                    B_{r+2n}(\gamma^{(r+2n)}(t_{r+2n})).
\end{align*}

Our main results are given in the following two theorems.  
\begin{theorem}
\label{thm-Duhamel-iterate-bd-0-dim1}
Assume that $d=1$ and $t_r\in[0,T]$. The estimate
\eqn
	\Big\| \, S^{(r, \alpha)} \int_{0}^{t_r} ... \int_{0}^{t_{r+2n-2}}
      J^r(\ut_{r+2n}) \; dt_{r+2} ... dt_{r+2n} \, \Big\|_{L^2(\R^{dr}\times\R^{dr})}
	\, < \, C^r \, (C_0 T)^n
\eeqn
holds for $\frac{1}{2} < \alpha \leq 1$,  
and for constants $C,C_0$ independent of $r$ and $T$. 
\end{theorem}

\begin{theorem}
\label{thm-Duhamel-iterate-bd-0}
Assume that $d=2$ and $t_r\in[0,T]$. The estimate
\eqn
	\Big\| \, S^{(r, \alpha)} \int_{0}^{t_r} ... \int_{0}^{t_{r+2n-2}}
      J^r(\ut_{r+2n}) \; dt_{r+2} ... dt_{r+2n} \, \Big\|_{L^2(\R^{dr}\times\R^{dr})}
	\, < \, C^r \, (C_0 T)^n
\eeqn
holds for $\frac{5}{6} < \alpha < 1$, and 
for constants $C,C_0$ independent of $r$ and $T$. 
\end{theorem}

Theorems \ref{thm-Duhamel-iterate-bd-0-dim1} and 
\ref{thm-Duhamel-iterate-bd-0} 
 imply that for sufficiently small $T$, 
\eqn
\Big\| \, S^{(r, \alpha)} \int_{0}^{t_r} ... \int_{0}^{t_{r+2n-2}}
      J^r(\ut_{r+2n}) \; dt_{r+2} ... dt_{r+2n} \, \Big\|_{L^2(\R^{dr}\times\R^{dr})}
	\, \rightarrow \, 0
\eeqn
as $n\rightarrow\infty$. Since $n$ is arbitrary, we conclude that
$\gamma^{(r)}(t_r, \cdot) =0$, given the initial condition $\gamma^{(r)}(0, \cdot) =0$. 
This establishes the uniqueness of $\gamma^{(r)}(t_r, \cdot)$,
and since $r$ is arbitrary, we conclude that the solution of the infinite
hierarchy is unique.

First, we shall present a proof of Theorem \ref{thm-Duhamel-iterate-bd-0-dim1},
which is done via iterative applications of the spatial bound obtained in Theorem \ref{highregbound}
followed by the use of  the a priori spatial bound given in Theorem \ref{KMbound}. 
We thank the referee and Aynur Bulut
who independently observed that  Theorem \ref{thm-Duhamel-iterate-bd-0-dim1}
can be proved without using the combinatorial argument of Section \ref{sect-Duhamel-1}
(which was used in an earlier version of the manuscript).
\\

\noindent{\em Proof} {\it{of Theorem  \ref{thm-Duhamel-iterate-bd-0-dim1} (joint with Aynur Bulut}}).
Let us fix $\alpha$ such that $\frac{1}{2} < \alpha \leq 1$. 
Since $e^{i(t_r - t_{r+2}) \Delta_{\pm}^{(r)} }$ commutes with 
the operator $S^{(r,\alpha)}$ and  $e^{i(t_r - t_{r+2}) \Delta_{\pm}^{(r)} }$ is unitary 
we have: 

\begin{align} 
      \Big\| \, & S^{(r, \alpha)} \int_{0}^{t_r} ... \int_{0}^{t_{r+2n-2}}
                  J^r(\ut_{r+2n}) \; dt_{r+2} ... dt_{r+2n} \, \Big\|_{L^2(\R^{r}\times\R^{r})} 
                   \nonumber \\
      & \leq   \int_{0}^{t_r} ... \int_{0}^{t_{r+2n-2}}  
                  \| S^{(r, \alpha)} e^{i(t_r - t_{r+2}) \Delta_{\pm}^{(r)} }
                  B_{r+2} e^{ i(t_{r+2} - t_{r+4})\Delta_{\pm}^{(r+2)} } B_{r+4} ...  
                  \nonumber \\
      &          \quad \quad \quad \quad e^{ i(t_{r+2(n-1)} - t_{r+2n})\Delta_{\pm}^{(r+2(n-1))} } B_{r+2n}(\gamma^{(r+2n)}(t_{r+2n})) \|_{ L^2(\R^{r} \times \R^{r})}   
                  \; dt_{r+2} ... dt_{r+2n} 
                  \nonumber\\
      & \leq   \int_{0}^{t_r} ... \int_{0}^{t_{r+2n-2}}  
                  \| S^{(r, \alpha)} B_{r+2} e^{ i(t_{r+2} - t_{r+4})\Delta_{\pm}^{(r+2)} } B_{r+4} ... e^{ i(t_{r+2(n-1)} - t_{r+2n})\Delta_{\pm}^{(r+2(n-1))} } 
                  \nonumber \\
      &          \quad \quad \quad \quad B_{r+2n}(\gamma^{(r+2n)}(t_{r+2n})) \|_{ L^2(\R^{r} \times \R^{r})}   
                  \; dt_{r+2} ... dt_{r+2n} 
                  \nonumber\\
       & \leq   r  \; \int_{0}^{t_r} ... \int_{0}^{t_{r+2n-2}}  
                  \| S^{(r+2, \alpha)} e^{ i(t_{r+2} - t_{r+4})\Delta_{\pm}^{(r+2)} } B_{r+4} ... e^{ i(t_{r+2(n-1)} - t_{r+2n})\Delta_{\pm}^{(r+2(n-1))} } 
                  \nonumber \\
      &          \quad \quad \quad \quad B_{r+2n}(\gamma^{(r+2n)}(t_{r+2n})) \|_{ L^2(\R^{r+2} \times \R^{r+2})}   
                  \; dt_{r+2} ... dt_{r+2n} 
                  \label{dim1-iterate1} \\
      & ...      \nonumber \\
      & \leq   r(r+2)\cdots (r+2n-4)  \; \int_{0}^{t_r} ... \int_{0}^{t_{r+2n-2}}  \| S^{(r+2n-2,\alpha)} e^{ i(t_{r+2(n-1)} - t_{r+2n})\Delta_{\pm}^{(r+2(n-1))} }
                  \nonumber \\
      &          \quad \quad \quad \quad  
                  B_{r+2n}(\gamma^{(r+2n)}(t_{r+2n})) \|_{ L^2(\R^{r+2n-2} \times \R^{r+2n-2)})}   
                  \; dt_{r+2} ... dt_{r+2n} 
                  \label{dim1-lastiterate} \\ 
      & \leq   r(r+2) \cdots (r+2n-4)(r+2n-2) \; \int_{0}^{t_r} ... \int_{0}^{t_{r+2n-2}}  
                  C^{r+2n}  \; dt_{r+2} ... dt_{r+2n} 
                  \label{dim1-KMbound} \\ 
      & \leq C^{r+2n} \; 2^n \;  \frac{ (\lceil\frac{r}{2}\rceil  - 1 + n )\, ! }{ (\lceil\frac{r}{2}\rceil - 1) \, ! } \;  \frac{t_r^n}{n!} 
                \label{dim1-aftercomb}  \\
      & = C^{r+2n} \; 2^n \; {{\lceil\frac{r}{2}\rceil  - 1 +n }\choose {n}} \; t_r^n 
	\\
      & \leq C^{r+2n} \; 2^n \; 2^{\lceil\frac{r}{2}\rceil  - 1 + n} \; t_r^n 
               \nonumber \\
      & \leq C^r (C_0 T)^n, 
               \label{dim1-atlast}     
\end{align} 
where to obtain \eqref{dim1-iterate1} we applied Theorem \ref{highregbound} and consequently continued iterative applications
of this bound in order to obtain \eqref{dim1-lastiterate}. Then to obtain \eqref{dim1-KMbound} we use the fact that 
 $e^{ i(t_{r+2(n-1)} - t_{r+2n})\Delta_{\pm}^{(r+2(n-1))} }$ commutes with  $S^{(r+2n-2,\alpha)}$, 
the unitarity of $e^{ i(t_{r+2(n-1)} - t_{r+2n})\Delta_{\pm}^{(r+2(n-1))} }$ and  
an application of the spatial a-priori bound stated in Theorem \ref{KMbound}. Finally, \eqref{dim1-aftercomb} was obtained by integrating and using the following simple observation: 
\begin{align*}
r(r+2)\cdots...(r+2n-2) 
& = 2^n \; \frac{r}{2}(\frac{r}{2}+1) \cdots(\frac{r}{2}+n-1) \\
& < 2^n \; \lceil\frac{r}{2} \; \rceil (\lceil\frac{r}{2}\rceil +1) \cdots(\lceil\frac{r}{2}\rceil + n - 1) \\
& = 2^n \; \frac{ (\lceil\frac{r}{2}\rceil  - 1 + n )\, ! }{ (\lceil\frac{r}{2}\rceil - 1) \, ! }.
\end{align*} 
Hence the theorem is proved. 
\qed

The proof of Theorem \ref{thm-Duhamel-iterate-bd-0} in the case $d=2$ will occupy sections \ref{sect-Duhamel-1} and 
\ref{sect-uniqproof-1}.

\section{Combinatorics of contractions} 
\label{sect-Duhamel-1}

In this section, we organize the Duhamel expansion with respect to the 
individual terms in the operators $B_{r+2\ell}$. 
This is obtained from an extension of the method of Klainerman-Machedon 
introduced in \cite{klma}.

Recalling that $B_{k+2} = \sum_{j=1}^k B_{j;k+1,k+2}$ 
we can rewrite  $J^r(\ut_{r+2n})$ as
\eqn 
J^r(\ut_{r+2n}) = \sum_{\mu \in M}  J^r(\ut_{r+2n};\mu),
\eeqn
where 
\begin{align*} 
J^r(\ut_{r+2n};\mu) = 
& e^{ i(t_r - t_{r+2})\Delta_{\pm}^{(r)} } 
  B_{\mu(r+1);r+1,r+2} e^{ i(t_{r+2} - t_{r+4})\Delta_{\pm}^{(r+2)} } 
  ...\\ 
& e^{ i(t_{r+2(n-1)} - t_{r+2n})\Delta_{\pm}^{(r+2(n-1))} } 
  B_{\mu(r+2n-1);r+2n-1,r+2n}(\gamma^{(r+2n)}(t_{r+2n})),
\end{align*} 
and $\mu$ is a map from $\{r+1,r+2,...,r+2n-1\}$ to 
$\{1, 2, 3, ... ,r+2n-2\}$ such that $\mu(2) = 1$ and $\mu(j) < j$ for all 
$j$. Here $M$ denotes the set of all such mappings $\mu$.  

We observe that such a mapping $\mu$ can be represented by highlighting 
one nonzero entry in each column of the $(r+2n-2) \times n$ matrix: 
\eqn \label{matrixB}
\left[ 
\begin{array}{cccc}
{\bf{B_{1;r+1,r+2}}} & B_{1;r+3,r+4}        & ... & {\bf{B_{1; r+2n-1, r+2n}}} \\
...                                & {\bf{B_{2;r+3,r+4}}} & ... & ... \\
...                                & ...                  & ... & ... \\
B_{r; r+1, r+2}          & B_{r;r+3, r+4}       & ... & ... \\
0                                 & B_{r+1;r+3,r+4}      & ... & ... \\
0                                 & B_{r+2;r+3,r+4}      & ... & ... \\
...                                & 0                    & ... & ... \\
...                                & ...                  & ... & ... \\
0                                 & 0                    & ... & B_{r+2n-2;r+2n-1,r+2n} 
\end{array} 
\right].
\eeqn

Since we can 
rewrite \eqref{iterates} as
\eqn \label{muJ}
\gamma^{(r)}(t_r, \cdot) = 
\int_{0}^{t_r} ... \int_{0}^{t_{r+2n-2}}
\sum_{\mu \in M} J^r(\ut_{r+2n},\mu) \; dt_{r+2} ... dt_{r+2n},
\eeqn
the integrals of the following type are of interest to us:  
\eqn \label{int-musig}
I(\mu, \sigma) = \int_{ t_{r} \geq t_{\sigma(r+2)} \geq ... \geq t_{\sigma(r+2n)} } 
J^r(\ut_{r+2n},\mu) \; dt_{r+2} ... dt_{r+2n},
\eeqn
where $\sigma$ is a permutation of $\{r+2,r+4,...,r+2n\}$.
We would like to associate to such an integral a matrix, which will
help us visualize $B_{\mu(r+2j-1);r+2j-1,r+2j}$'s as well as $\sigma$ at the same time. 
More precisely, to $I(\mu, \sigma)$ we associate the matrix 
$$ 
\left[ 
\begin{array}{cccc}
t_{\sigma^{-1}(r+2)} & t_{\sigma^{-1}(r+4)} & ... & t_{\sigma^{-1}(r+2n)} \\
{\bf{B_{1;r+1,r+2}}} & B_{1;r+3,r+4}        & ... & {\bf{B_{1; r+2n-1, r+2n}}} \\
...                  & {\bf{B_{2;r+3,r+4}}} & ... & ... \\
...                  & ...                  & ... & ... \\
B_{r; r+1, r+2}      & B_{r;r+3, r+4}       & ... & ... \\
0                    & B_{r+1;r+3,r+4}      & ... & ... \\
0                    & B_{r+2;r+3,r+4}      & ... & ... \\
...                  & 0                    & ... & ... \\
...                  & ...                  & ... & ... \\
0                    & 0                    & ... & B_{r+2n-2;r+2n-1,r+2n} 
\end{array} 
\right]$$
whose columns are labeled $1$ through $n$ and whose rows are labeled
$0, 1, ... r+2n-2$.

As in \cite{klma} we introduce a board game on the set of such matrices. 
In particular, the following move shall be called an ``acceptable move'':
If $\mu(r+2j+1) < \mu(r+2j-1)$, the player as allowed to
do the following four changes at the same time: 
\begin{itemize} 
\item exchange the highlightened entries in columns $j$ and $j+1$,  
\item exchange the highlightened entries in rows $r+2j-1$ and $r+2j+1$, 
\item exchange the highlightened entries in rows $r+2j$ and $r+2j+2$, 
\item exchange $t_{\sigma^{-1}(r+2j)}$ and $t_{\sigma^{-1}(r+2j+2)}$.
\end{itemize}

As in \cite{klma}, the importance of this game is visible from  
the following lemma: 
\begin{lemma} \label{transformI}
If $(\mu,\sigma)$ is transformed into $(\mu',\sigma')$ by an acceptable move, 
then $I(\mu,\sigma) = I(\mu',\sigma')$. 
\end{lemma} 

\prf 
We modify the proof of Lemma 3.1 in \cite{klma} accordingly. 
Let us start by fixing an integer $j\geq 3$. Then select two integers $i$ and $l$ such that 
$i < l < j < j+1$. 

Suppose $I(\mu, \sigma)$  and $I(\mu',\sigma')$ are as follows 
\begin{align} 
                       I(\mu, \sigma)
                       & = \int_{ t_{r} \geq ... \geq t_{\sigma(r+2j)} \geq t_{\sigma(r+2j+2)} \geq ... \geq t_{\sigma(r+2n)}\geq 0 } 
                              J^r(\ut_{r+2n},\mu) \; dt_{r+2} ... dt_{r+2n} \nonumber \\
                       & = \int_{ t_{r} \geq ... \geq t_{\sigma(r+2j)} \geq t_{\sigma(r+2j+2)} \geq ... \geq t_{\sigma(r+2n)}\geq 0 }        
                              ... e^{ i(t_{r+2j-2} - t_{r+2j})\Delta_{\pm}^{(r+2j-2)} } \nonumber \\
                       & \quad \quad B_{l;r+2j-1,r+2j}  e^{ i(t_{r+2j} - t_{r+2j+2})\Delta_{\pm}^{(r+2j)} }B_{i;r+2j+1,r+2j+2} \nonumber \\ 
                       & \quad \quad e^{ i(t_{r+2j+2} - t_{r+2j+4})\Delta_{\pm}^{(r+2j+2)} } (...) \; dt_{r+2} ... dt_{r+2n},\label{I}
\end{align} 

and 

\begin{align} 
                       I(\mu', \sigma')
                       & = \int_{ t_{r} \geq ... \geq t_{\sigma'(r+2j)} \geq t_{\sigma'(r+2j+2)} \geq ... \geq t_{\sigma'(r+2n)}\geq 0 } 
                              J^r(\ut_{r+2n},\mu') \; dt_{r+2} ... dt_{r+2n} \nonumber \\
                       & = \int_{ t_{r} \geq ... \geq t_{\sigma'(r+2j)} \geq t_{\sigma'(r+2j+2)} \geq ... \geq t_{\sigma'(r+2n)}\geq 0 }        
                              ... e^{ i(t_{r+2j-2} - t_{r+2j})\Delta_{\pm}^{(r+2j-2)} } \nonumber \\
                       & \quad \quad B_{i;r+2j-1,r+2j}  e^{ i(t_{r+2j} - t_{r+2j+2})\Delta_{\pm}^{(r+2j)} }B_{l;r+2j+1,r+2j+2} \nonumber \\ 
                       & \quad \quad e^{ i(t_{r+2j+2} - t_{r+2j+4})\Delta_{\pm}^{(r+2j+2)} } (...)' \; dt_{r+2} ... dt_{r+2n}.\label{Iprime}
\end{align} 
Here $...$ in \eqref{I} and \eqref{Iprime} coincide. On the other hand any $B_{r+2j-1;s,s+1}$ in $(...)$ of \eqref{I} becomes 
$B_{r+2j+1;s,s+1}$ in $(...)'$ of \eqref{Iprime} and any $B_{r+2j+1;s,s+1}$ in $(...)$ of \eqref{I} becomes 
$B_{r+2j-1;s,s+1}$ in $(...)'$ of \eqref{Iprime}. Also any $B_{r+2j;s,s+1}$ in $(...)$ of \eqref{I} becomes 
$B_{r+2j+2;s,s+1}$ in $(...)'$ of \eqref{Iprime} and any $B_{r+2j+2;s,s+1}$ in $(...)$ of \eqref{I} becomes 
$B_{r+2j;s,s+1}$ in $(...)'$ of \eqref{Iprime}. 

We shall prove that 
\eqn \label{Iequality} 
I(\mu,\sigma) = I(\mu',\sigma').
\eeqn

As in \cite{klma}, we introduce $P$ and $\tilde{P}$ which are, in our context, defined as follows: 
\begin{align*} 
                        P & = B_{l;r+2j-1,r+2j}  e^{ i(t_{r+2j} - t_{r+2j+2})\Delta_{\pm}^{(r+2j)} }B_{i;r+2j+1,r+2j+2},\\
                        \tilde{P} & = B_{i;r+2j+1,r+2j+2}  e^{ -i(t_{r+2j} - t_{r+2j+2})\tilde{\Delta}_{\pm}^{(r+2j)} }B_{l;r+2j-1,r+2j},                           
\end{align*}
where 
$$ 
     \tilde{\Delta}_{\pm}^{(r+2j)} = \Delta_{\pm}^{(r+2j)} - \Delta_{\pm,x_{r+2j}} - \Delta_{\pm,x_{r+2j-1}}  + 
                                                         \Delta_{\pm,x_{r+2j+1}} + \Delta_{\pm,x_{r+2j+2}}. 
$$                                                         
 
First, let us prove that 
\eqn
	\lefteqn{
	e^{ i(t_{r+2j-2} - t_{r+2j})\Delta_{\pm}^{(r+2j-2)} }  P e^{ i(t_{r+2j+2} - t_{r+2j+4})\Delta_{\pm}^{(r+2j+2)} }
	} 
	\nonumber\\
     && = \,  e^{ i(t_{r+2j-2} - t_{r+2j+2})\Delta_{\pm}^{(r+2j-2)} } 
     \tilde{P} e^{ i(t_{r+2j} - t_{r+2j+4})\Delta_{\pm}^{(r+2j+2)} }.\label{commute}
\eeqn        
In order to do that we observe that 
$$ 
     \Delta_{\pm}^{(r+2j)} = \Delta_{\pm,x_i} + (\Delta_{\pm}^{(r+2j)} - \Delta_{\pm,x_i}).
$$ 
Hence the factor $e^{ i(t_{r+2j} - t_{r+2j+2})\Delta_{\pm}^{(r+2j)} }$ 
appearing in the definition of $P$ can be rewritten as 
\eqn
	\lefteqn{
	e^{ i(t_{r+2j} - t_{r+2j+2})\Delta_{\pm}^{(r+2j)} } 
	}
	\nonumber\\
     &&= \,  e^{ i(t_{r+2j} - t_{r+2j+2})\Delta_{\pm,x_i} }
      e^{ i(t_{r+2j} - t_{r+2j+2})(\Delta_{\pm}^{(r+2j)} - \Delta_{\pm,x_i}) }, 
\eeqn      
which in turn allows us to see (after two basic commutations) that $P$ equals to: 
\begin{align}  
             P = & e^{ i(t_{r+2j} - t_{r+2j+2})\Delta_{\pm,x_i} }  
                   B_{l;r+2j-1,r+2j} B_{i;r+2j+1,r+2j+2} \nonumber \\ 
                 & e^{ i(t_{r+2j} - t_{r+2j+2})(\Delta_{\pm}^{(r+2j)} - \Delta_{\pm,x_i}) }. \label{Paftercomm} 
\end{align}  
Therefore using \eqref{Paftercomm}, the LHS of \eqref{commute} can be rewritten as 
\begin{align} 
             & e^{ i(t_{r+2j-2} - t_{r+2j})\Delta_{\pm}^{(r+2j-2)} } 
               P e^{ i(t_{r+2j+2} - t_{r+2j+4})\Delta_{\pm}^{(r+2j+2)} } \nonumber \\
     = \quad & e^{ i(t_{r+2j-2} - t_{r+2j})\Delta_{\pm}^{(r+2j-2)} } 
               e^{ i(t_{r+2j} - t_{r+2j+2})\Delta_{\pm,x_i} } \nonumber \\
             & B_{l;r+2j-1,r+2j} B_{i;r+2j+1,r+2j+2} \nonumber \\ 
             & e^{ i(t_{r+2j} - t_{r+2j+2})(\Delta_{\pm}^{(r+2j)} - \Delta_{\pm,x_i}) } 
               e^{ i(t_{r+2j+2} - t_{r+2j+4})\Delta_{\pm}^{(r+2j+2)} } \nonumber \\ 
     = \quad & e^{ i(t_{r+2j-2} - t_{r+2j})\Delta_{\pm}^{(r+2j-2)} } 
               e^{ i(t_{r+2j} - t_{r+2j+2})\Delta_{\pm,x_i} } \nonumber \\
             & B_{l;r+2j-1,r+2j} B_{i;r+2j+1,r+2j+2} \nonumber \\
             & e^{ i(t_{r+2j+2} - t_{r+2j+4})(\Delta_{\pm,x_i} + \Delta_{\pm,r+2j+1} + 
                \Delta_{\pm,r+2j+2}) } \nonumber\\
             & \quad \quad \quad \quad \quad \quad
              e^{ i(t_{r+2j} - t_{r+2j+4}) (\Delta_{\pm,x_1} + ... + \hat{\Delta}_{\pm,x_i} + ... + \Delta_{\pm,r+2j}) }, 
             \label{lhs-commute}
\end{align} 
where $\hat{\Delta}_{\pm,x_i}$ denotes that the term ${\Delta}_{\pm,x_i}$ is missing.\\ 
On the other hand, we can rewrite $\tilde{\Delta}_{\pm}^{(r+2j)}$ as
\begin{align*} 
              \tilde{\Delta}_{\pm}^{(r+2j)} 
              & = \Delta_{\pm}^{(r+2j)} - \Delta_{\pm,x_{r+2j}} - \Delta_{\pm,x_{r+2j-1}}  
                  + \Delta_{\pm,x_{r+2j+1}} + \Delta_{\pm,x_{r+2j+2}} \\
              & = \Delta_{\pm}^{(r+2j-2)} 
                  + \Delta_{\pm,x_{r+2j+1}} + \Delta_{\pm,x_{r+2j+2}} \\
              & = (\Delta_{\pm}^{(r+2j-2)} - \Delta_{\pm,x_i}) 
                  + (\Delta_{\pm,x_i} + \Delta_{\pm,x_{r+2j+1}} + \Delta_{\pm,x_{r+2j+2}}). 
\end{align*}
Hence the factor $e^{ -i(t_{r+2j} - t_{r+2j+2})\tilde{\Delta}_{\pm}^{(r+2j)} }$ appearing in the definition of $\tilde{P}$ can be rewritten as: 
\begin{align*}
              & e^{ -i(t_{r+2j} - t_{r+2j+2})\tilde{\Delta}_{\pm}^{(r+2j)} } \\
              = \quad & e^{ -i(t_{r+2j} - t_{r+2j+2})(\Delta_{\pm}^{(r+2j-2)} - \Delta_{\pm,x_i}) } 
    e^{ -i(t_{r+2j} - t_{r+2j+2})(\Delta_{\pm,x_i} + \Delta_{\pm,x_{r+2j+1}} + \Delta_{\pm,x_{r+2j+2}}) },
\end{align*}
which in turn implies that (after two basic commutations) $\tilde{P}$ equals 
\begin{align} 
             \tilde{P} = & e^{ -i(t_{r+2j} - t_{r+2j+2})(\Delta_{\pm}^{(r+2j-2)} - \Delta_{\pm,x_i}) }    
                           B_{i;r+2j+1,r+2j+2} B_{l;r+2j-1,r+2j}\nonumber \\
                         & e^{ -i(t_{r+2j} - t_{r+2j+2})(\Delta_{\pm,x_i} + \Delta_{\pm,x_{r+2j+1}} + \Delta_{\pm,x_{r+2j+2}}) }.\label{tilPaftercomm}         
\end{align}
Thus using \eqref{tilPaftercomm}, the RHS of \eqref{commute} can be written as 
\begin{align} 
             & e^{ i(t_{r+2j-2} - t_{r+2j+2})\Delta_{\pm}^{(r+2j-2)} } 
               \tilde{P} e^{ i(t_{r+2j} - t_{r+2j+4})\Delta_{\pm}^{(r+2j+2)} } \nonumber \\
     = \quad & e^{ i(t_{r+2j-2} - t_{r+2j+2})\Delta_{\pm}^{(r+2j-2)} } 
               e^{ -i(t_{r+2j} - t_{r+2j+2})(\Delta_{\pm}^{(r+2j-2)} - \Delta_{\pm,x_i}) } \nonumber\\    
             & B_{i;r+2j+1,r+2j+2} B_{l;r+2j-1,r+2j}\nonumber \\
             & e^{ -i(t_{r+2j} - t_{r+2j+2})(\Delta_{\pm,x_i} + \Delta_{\pm,x_{r+2j+1}} + \Delta_{\pm,x_{r+2j+2}}) }  
               e^{ i(t_{r+2j} - t_{r+2j+4})\Delta_{\pm}^{(r+2j+2)} } \nonumber \\
     = \quad & e^{ i(t_{r+2j-2} - t_{r+2j})\Delta_{\pm}^{(r+2j-2)} } 
               e^{ i(t_{r+2j} - t_{r+2j+2})\Delta_{\pm,x_i} } \nonumber \\
             & B_{i;r+2j+1,r+2j+2} B_{l;r+2j-1,r+2j}\nonumber \\
             & e^{ i(t_{r+2j+2} - t_{r+2j+4})(\Delta_{\pm,x_i} + \Delta_{\pm,r+2j+1} + 
                \Delta_{\pm,r+2j+2}) } \nonumber\\
             & \quad \quad \quad \quad \quad \quad
                e^{ i(t_{r+2j} - t_{r+2j+4}) (\Delta_{\pm,x_1} + ... + \hat{\Delta}_{\pm,x_i} + ... + \Delta_{\pm,r+2j}) }. \label{rhs-commute}
\end{align}                            
We combine \eqref{lhs-commute} and \eqref{rhs-commute} to obtain \eqref{commute}.

Now we are ready to prove \eqref{Iequality}.  We observe that thanks to the symmetry the value of $I(\mu,\sigma)$ does not
change if in \eqref{I} we perform the following two exchanges in the arguments of $\gamma^{(r+2n)}$ only: 
\begin{itemize}
\item exchange $(x_{r+2j-1},x'_{r+2j-1})$ with $(x_{r+2j+1},x'_{r+2j+1})$ 
\item exchange $(x_{r+2j},x'_{r+2j})$ with $(x_{r+2j+2},x'_{r+2j+2})$. 
\end{itemize} 
After these two exchanges we use \eqref{commute} and the definition of $\tilde{P}$ to rewrite \eqref{I} as:
\begin{align} 
                       & I(\mu, \sigma) \nonumber \\ 
                    = & \int_{ t_{r} \geq ... \geq t_{\sigma(r+2j)} \geq t_{\sigma(r+2j+2)} \geq ... \geq t_{\sigma(r+2n)} \geq 0} ...\nonumber \\       
                       &  \quad \quad e^{ i(t_{r+2j-2} - t_{r+2j})\Delta_{\pm}^{(r+2j-2)} } 
                            P e^{ i(t_{r+2j+2} - t_{r+2j+4})\Delta_{\pm}^{(r+2j+2)} } (...)' \; dt_{r+2} ... dt_{r+2n} \nonumber \\
                    = & \int_{ t_{r} \geq ... \geq t_{\sigma(r+2j)} \geq t_{\sigma(r+2j+2)} \geq ... \geq t_{\sigma(r+2n)}\geq 0}...\nonumber \\        
                       &  \quad \quad e^{ i(t_{r+2j-2} - t_{r+2j+2})\Delta_{\pm}^{(r+2j-2)} }  
                            \tilde{P} e^{ i(t_{r+2j} - t_{r+2j+4})\Delta_{\pm}^{(r+2j+2)} }(...)' \; dt_{r+2} ... dt_{r+2n} \nonumber \\
                    = & \int_{ t_{r} \geq ... \geq t_{\sigma(r+2j)} \geq t_{\sigma(r+2j+2)} \geq ... \geq t_{\sigma(r+2n)}\geq 0 } 
                           \int_{\R^{d(r+2n+2)}}
                              ... e^{ i(t_{r+2j-2} - t_{r+2j+2})\Delta_{\pm}^{(r+2j-2)} }\nonumber \\ 
                       &    \quad \quad \delta_{i;r+2j+1,r+2j+2}  e^{ -i(t_{r+2j} - t_{r+2j+2})\tilde{\Delta}_{\pm}^{(r+2j)} }\delta_{l;r+2j-1,r+2j}\nonumber\\    
                       &    \quad \quad e^{ i(t_{r+2j} - t_{r+2j+4})\Delta_{\pm}^{(r+2j+2)} }(...)' \; dt_{r+2} ... dt_{r+2n},  \label{usingdel}
\end{align}                       
where $\delta_{j;k+1,k+2}$ denotes the kernel of the operator $B_{i;k+1,k+2}$ i.e.
\eqn 
	\delta_{j;k+1,k+2}  
	& = & \delta(x_j-x_{k+1})\delta(x_j-x'_{k+1})\delta(x_j-x_{k+2})\delta(x_j-x'_{k+2}) 
	\\
	&&\quad\quad\quad 
	- \delta(x'_j-x_{k+1})\delta(x'_j-x'_{k+1})\delta(x'_j-x_{k+2})\delta(x'_j-x'_{k+2}) \, . 
	\nonumber
\eeqn                                        
Now in \eqref{usingdel} we perform the change of variables that exchanges
$$
	(t_{r+2j-1}, x_{r+2j-1},x'_{r+2j-1})
	\; \; \; \; \text{and} \; \; \; \; 
	(t_{r+2j+1}, x_{r+2j+1},x'_{r+2j+1})
$$  
as well as 
$$
	(t_{r+2j}, x_{r+2j},x'_{r+2j})
	\; \; \; \; \text{and} \; \; \; \; 
	(t_{r+2j+2},x_{r+2j+2},x'_{r+2j+2}) \,.
$$ 
Under the same change of variables 
$\tilde{\Delta}^{(r+2j)}$ which is equal to 
\begin{align*} 
                        \tilde{\Delta}^{(r+2j)} & = \Delta_{\pm}^{(r+2j)} - \Delta_{\pm,x_{r+2j}} - \Delta_{\pm,x_{r+2j-1}}  + 
                                                                    \Delta_{\pm,x_{r+2j+1}} + \Delta_{\pm,x_{r+2j+2}} \\    
                                                            & = \Delta_{\pm}^{(r+2j-2)}  + \Delta_{\pm,x_{r+2j-1}} + \Delta_{\pm,x_{r+2j}}                                                         
                                                                   \\
                                                                   &\quad\quad\quad\quad\quad\quad
                                                                   - \Delta_{\pm,x_{r+2j}} - \Delta_{\pm,x_{r+2j-1}}  
                                                                   + \Delta_{\pm,x_{r+2j+1}} + \Delta_{\pm,x_{r+2j+2}}\\ 
                                                            & =  \Delta_{\pm}^{(r+2j-2)} + \Delta_{\pm,x_{r+2j+1}} + \Delta_{\pm,x_{r+2j+2}} 
\end{align*} 
becomes                                                  
$ \Delta_{\pm}^{(r+2j-2)} + \Delta_{\pm,x_{r+2j-1}} + \Delta_{\pm,x_{r+2j}}$  that equals  
$\Delta_{\pm}^{(r+2j)}$. Therefore, after we perform this change of variables in \eqref{usingdel}, we obtain 
\eqn
	I(\mu, \sigma)
	& = & \int_{ t_{r} \geq \cdots \geq t_{\sigma'(r+2j+2)} \geq t_{\sigma'(r+2j)} \geq \cdots \geq t_{\sigma(r+2n)} }        
	\cdots e^{ i(t_{r+2j-2} - t_{r+2j})\Delta_{\pm}^{(r+2j-2)} }
	\nonumber\\ 
	&&\quad \quad B_{i;r+2j-1,r+2j}  e^{ -i(t_{r+2j+2} - t_{r+2j})\Delta_{\pm}^{(r+2j)} }B_{l;r+2j+1,r+2j+2}
	\nonumber\\    
	&&    \quad \quad e^{ i(t_{r+2j+2} - t_{r+2j+4})\Delta_{\pm}^{(r+2j+2)} }(\cdots)' \; dt_{r+2} \cdots dt_{r+2n} 
	\nonumber\\  
	&  = & I(\mu',\sigma'),
\eeqn     
where $\sigma' = (r+2j, r+2j+2) \circ \sigma$. Here $(a,b)$ denotes the permutation which reverses $a$ and $b$.  
Hence \eqref{Iequality} is proved. 
\endprf

Let us consider the set $N$ of those matrices in $M$ which are in so-called ``upper echelon" form. 
Here, as in \cite{klma}, we say that a matrix of the type \eqref{matrixB} is in upper echelon form if each highlighted entry in a row is 
to the left of each highlighted entry in a lower row. For example, the following matrix is in upper echelon form  (with $r=1$ and $n=3$):
$$ 
\left[ 
\begin{array}{ccc}
{\bf{B_{1;2,3}}} & B_{1;4,5}  &  B_{1;6,7} \\
0 & {\bf{B_{2; 4,5}}} & B_{2;6,7} \\
0 &  B_{3;4,5} & B_{3;6,7}\\
0 & 0 & {\bf{B_{4;6,7}}}\\
0 & 0 & B_{5;6,7}
\end{array} 
\right].
$$
In the same way as in Lemma 3.2 in \cite{klma} one can prove that in our context: 
\begin{lemma} 
For each matrix in $M$ there is a finite number of acceptable moves that transforms the matrix 
into upper echelon form. 
\end{lemma} 

Let $C_{r,n}$ denote the number of upper echelon matrices of the size $(r +2n - 2) \times n$. The following lemma gives an upper bound on 
$C_{r,n}$. 
\begin{lemma} 
The following holds: 
$$C_{r,n}  \leq 2^{r+3n-2}.$$ 
\end{lemma}

\prf
As in \cite{klma} the proof proceeds in two steps. 

\begin{enumerate} 
\item[\bf{Step 1}]
First, we bring all highlighted entries to the first row. In such a way the first row
is partitioned into subsets that consist of elements that were originally in the same row.
Let us denote by $P_n$ the number of possible partitions of the first row into these subsets. 
Then 
$$P_n \leq 2^n,$$ 
ax explained in \cite{klma}.  One can see this by first observing that 
\eqn \label{Pnequal}
P_n = 1 + P_1 + \dots + P_{n-1},
\eeqn
 which in turn can be verified by counting the number of the elements in the last subset. More precisely, if the last subset has $0$ elements that gives exactly one contribution toward $P_n$. In general, if the last subset has $k$ elements, then the rest of $n-k$ elements of the first 
row can be partitioned into $P_{n-k}$ ways. Hence \eqref{Pnequal} follows. 

\item[\bf{Step 2}]
Now we reassemble the matrix obtained in the previous step by lowering the first subset into the first used row, the second subset into 
the second used row etc. If a given partition of the first row has exactly $i$ subsets, then these subsets can be lowered in an order preserving way to the available $r+2n-2$ rows in $ \binom{r+2n-2}{i}$ ways. 
\end{enumerate}

Now we combine Steps 1 and 2 to conclude 
$$C_{r,n} \leq P_n \sum_{i=1}^{n} \binom{r+2n-2}{i}  \leq 2^{r+3n-2}.$$
\endprf

Let $\mu_{es}$ be a matrix in $N$.   
We write $\mu \sim \mu_{es}$ if $\mu$ can be transformed into $\mu_{es}$ in finitely many acceptable moves.  
It can be seen that:
\begin{theorem}
\label{thm-Duhamel-iterate-bd-2} 
Suppose $\mu_{es} \in N$. Then there exists a subset of $[0,t_r]^n$, denoted by $D$, such that 
$$
                              \sum_{\mu \sim \mu_{es}} \int_{0}^{t_r} ... \int_{0}^{t_{r+2n-2}}
			  J^r(\ut_{r+2n},\mu) \; dt_{r+2} ... dt_{r+2n} 
                             = \int_{D}  J^r(\ut_{r+2n},\mu) \; dt_{r+2} ... dt_{r+2n}. 
$$                             
\end{theorem} 

\prf Here we give an outline of the proof, which is analogous to the proof 
of a similar result stated in Theorem 3.4 in \cite{klma}. 

We consider the integral 
$$ 
   I(\mu, id) = 
   \int_{0}^{t_r} ... \int_{0}^{t_{r+2n-2}}
   J^r(\ut_{r+2n},\mu) \; dt_{r+2} ... dt_{r+2n}. 
$$ 
and perform finitely many acceptable moves on the corresponding matrix
determined by $(\mu,id)$ until we transform it to the special upper echelon 
matrix associated with $(\mu_{es},\sigma)$. Then Lemma \ref{transformI} guarantees that  
$$ I(\mu, id) = I(\mu_{es}, \sigma).$$ 

As in \cite{klma}, if $(\mu_1, id)$ and $(\mu_2, id)$ with $\mu_1 \neq \mu_2$ 
produce the same echelon form $\mu_{es}$, then the corresponding permutations $\sigma_1$ 
and $\sigma_2$ must be distinct. Hence, to determine $D$, we need to identify all 
permutations $\sigma$ that occur in a connection with a given class of equivalence 
$\mu_{es}$. Then $D$ can be chosen to be the union of all 
$\{ t_r \geq t_{\sigma(r+2)} \geq t_{\sigma(r+4)} \geq ... \geq t_{\sigma(r+2n)}\}$. 
\endprf

\section{Proof of Theorem \ref{thm-Duhamel-iterate-bd-0} }
\label{sect-uniqproof-1}

In combination with Theorem \ref{thm-Duhamel-iterate-bd-2},
the following result immediately implies Theorem \ref{thm-Duhamel-iterate-bd-0}.

\begin{theorem}
\label{thm-Duhamel-iterate-bd-1}
Assume that $d = 2$ and $\frac56 < \alpha < 1$ and $t_r\in[0,T]$. The estimate
\eqn
	\Big\| \, S^{(r, \alpha)} \int_{D}  J^r(\ut_{r+2n},\mu) \; dt_{r+2} ... dt_{r+2n} \, \Big\|_{L^2(\R^{dr}\times\R^{dr})}
	\, < \, C^r \, (C_0 T)^n
\eeqn
holds for a constant $C_0$ independent of $r$ and $T$. 
\end{theorem}

\prf
The proof proceeds precisely in the same way as in \cite{klma,kiscst},
under the condition that $\frac56<\alpha<1$,
by using nested Duhamel formulas of section \ref{sect-Duhamel-1}, 
recursive applications of the space-time bounds given in Theorem \ref{thm-spacetime-bd-1}
and at the end by using the a priori spatial bound provided by Theorem \ref{KMbound}. 
\endprf

\appendix \section{Poincar\'e type inequality} 

Here we state and prove the following Poincar\'e type inequality. 

\begin{lemma} \label{lemma-poincare} 
Suppose that $h \in L^1({\R}^d)$ is a probability measure such that $\int_{{\R}^d} dx \; (1+x^2)^{1/2} h(x) < \infty$. 
Let $h_a(x) =  \frac{1}{a^d} h(\frac{x}{a})$. Then, for every $0 \leq \kappa < 1$, there exists $C > 0$ such that 
\begin{align} 
|\tr \; J^{(k)} & \left( h_a(x_j - x_{k+1}) h_a(x_j - x_{k+2}) - \delta(x_j - x_{k+1}) \delta(x_j - x_{k+2})  \right) \gamma^{(k+2)}| \nonumber \\
                 & \leq C a^{\kappa} \; |\!|\!| J^{(k)} |\!|\!| \; \tr |S_j S_{k+1} S_{k+2} \gamma^{(k+2)} S_{k+2} S_{k+1} S_j|, \label{ineq-poincare}
\end{align}                  
for all non-negative $\gamma^{(k+2)} \in {\mathcal L}_{k+2}^1$. 
\end{lemma} 

\prf We shall prove the lemma by modifying the proof of Lemma A.2 from \cite{kiscst}. As in \cite{kiscst}, we
demonstrate the argument for $k=1$ (the proof is analogous for $k>1$). 

We start by using the representation $\gamma^{(3)} = \sum_j \lambda_j \big|\, \phi_j  \, \ket \bra \, \phi_j \,\big|$, 
where $\phi_j \in L^2({\R}^{3d})$, for eigenvalues  $\lambda_j  \geq 0$. 
Therefore, 
\begin{align} 
& \tr \; J^{(1)} \left( h_a(x_1 - x_2) h_a(x_1 - x_3) - \delta(x_1 - x_2) \delta(x_1 - x_3)  \right) \gamma^{(3)} \nonumber \\
                    & =  \sum_{j} \lambda_j \, \bra \phi_j, J^{(1)} 
                    \left( h_a(x_1 - x_2) h_a(x_1 - x_3) - \delta(x_1 - x_2) \delta(x_1 - x_3)  
                    \right) \phi_j \ket \nonumber \\
                    & =  \sum_{j} \lambda_j \, \bra \psi_j, \left( h_a(x_1 - x_2) h_a(x_1 - x_3) - \delta(x_1 - x_2) \delta(x_1 - x_3)  \right) \phi_j \ket, \label{poinc-decomp}
\end{align}  
where $\psi_j = (J^{(1)} \otimes 1)\phi_j$. From Parseval's identity, we find
\begin{align} 
\bra \psi_j &, \left( h_a(x_1 - x_2) h_a(x_1 - x_3) 
	- \delta(x_1 - x_2) \delta(x_1 - x_3)  \right) \phi_j \ket \nonumber \\
                & = \int dp_1 \, dp_2 \, dp_3 \, dq_1 \, dq_2 \, dq_3 \,  
                {\overline{\widehat{\psi}_j (p_1, p_2, p_2)}} {\widehat{\phi}}_j(q_1, q_2, q_3) 
                \nonumber \\
                & \quad \quad \int dx \, dy \, h(x)\, h(y)  \; 
                \left( e^{i a x \cdot (p_2 - q_2)} e^{i a y (p_3-q_3)}  - 1\right) \nonumber \\
                &  \quad \quad \quad \quad  \delta(p_1 + p_2 + p_3  - q_1 - q_2 - q_3) \,.
                \label{poinc-parseval} 
\end{align} 
For arbitrary $0 < \kappa < 1$,  we have
\eqn
	| e^{i a x \cdot (p_2 - q_2)} e^{i a y (p_3-q_3)}  - 1 |
		& \leq & a^\kappa \, ( |x| \, |p_2-q_2| \, + \, |y| \, |p_3-q_3| )^\kappa
			\nonumber\\
		& \leq & a^\kappa \, ( |x|^\kappa \, |p_2-q_2|^\kappa 
		\, + \, |y|^\kappa \, |p_3-q_3|^\kappa ) \,.
\eeqn
Here we use $(a+b)^\kappa=((a^\kappa)^{\frac1\kappa}+(b^\kappa)^{\frac1\kappa})^\kappa
\leq ((a^\kappa + b^\kappa)^{\frac1\kappa})^\kappa= a^\kappa+b^\kappa$ for $a,b\geq0$, 
which holds whenever $\frac1\kappa\geq1$.
Thus, taking absolute values in \eqref{poinc-parseval}, and recalling that $\int h=1$,
we find  
\begin{align} 
| \bra \psi_j &, \left( h_a(x_1 - x_2) h_a(x_1 - x_3) - \delta(x_1 - x_2) \delta(x_1 - x_3)  \right) \phi_j \ket | \nonumber \\
                & \leq   a^{\kappa} \left( \int dx \, h(x) |x|^{\kappa} \right) \, 
	                   \int dp_1 \, dp_2 \, dp_3 \, dq_1 \, dq_2 \, dq_3 \, 
	                   (|p_2 -q_2|^\kappa \, + \, | p_3- q_3|^{\kappa}) \, \nonumber \\  
                & \quad \quad \quad \quad |{\widehat{\psi}}_j (p_1, p_2, p_2)| \, |{\widehat{\phi}}_j(q_1, q_2, q_3)| \,  \delta(p_1 + p_2 + p_2 - q_1 - q_2 - q_3).\label{poinc-afterabs} 
\end{align} 
Clearly, $|p_j - q_j|^{\kappa} \leq |p_j|^{\kappa} + |q_j|^{\kappa}$, for $j=2,3$,
and it suffices to show how to control one of the four terms thereby obtained,
for instance the one containing the factor $|p_2|^{\kappa}$. 
By applying Cauchy-Schwarz, we obtain: 
\begin{align} 
\int dp_1 & \, dp_2 \, dp_3  \, dq_1 \, dq_2 \, dq_3 \,  \delta(p_1 + p_2 + p_3  - q_1 - q_2 - q_3) \nonumber \\
                                       & \quad \quad \quad |{\widehat{\psi}}_j (p_1, p_2, p_2)| \,  |{\widehat{\phi}}_j(q_1, q_2, q_3)| \, |p_2|^{\kappa} \nonumber\\          
                                       & = \int dp_1 \, dp_2 \, dp_3 \, dq_1 \, dq_2 \, dq_3 \,  \delta(p_1 + p_2 + p_3  - q_1 - q_2 - q_3) \nonumber \\
                                       & \quad \quad  \quad \frac{\bra p_1 \ket \, \bra p_2 \ket \, \bra  p_3 \ket}{\bra q_1 \ket \, \bra q_2 \ket \,  \bra q_3 \ket} |{\widehat{\psi}}_j (p_1, p_2, p_2)|\,                             
                                                               \frac{\bra q_1 \ket \, \bra q_2 \ket \, \bra  q_3 \ket}{\bra p_1 \ket \, \bra p_2 \ket^{1-\kappa} \, \bra  p_3 \ket} |{\widehat{\phi}}_j (q_1, q_2, q_2)| \nonumber\\  
                                       &  \leq \epsilon \int dp_1 \, dp_2 \, dp_3 \, dq_1 \, dq_2 \, dq_3 \,  \delta(p_1 + p_2 + p_3  - q_1 - q_2 - q_3) \nonumber \\
                                       & \quad \quad  \quad \frac{\bra p_1 \ket^2 \, \bra p_2 \ket^2 \, \bra  p_3 \ket^2}{\bra q_1 \ket^2 \, \bra q_2 \ket^2 \, \bra  q_3 \ket^2} |{\widehat{\psi}}_j (p_1, p_2, p_2)|^2 \nonumber \\
                                       &  +  \epsilon^{-1} \int dp_1 \, dp_2 \, dp_3 \, dq_1 \, dq_2 \, dq_3 \,  \delta(p_1 + p_2 + p_3  - q_1 - q_2 - q_3) \nonumber \\
                                       & \quad \quad \quad \frac{\bra q_1 \ket^2 \, \bra q_2 \ket^2 \, \bra  q_3 \ket^2}{\bra p_1 \ket^2 \, \bra p_2 \ket^{2(1-\kappa)} \, \bra  p_3 \ket^2} |{\widehat{\phi}}_j (q_1, q_2, q_2)|^2 \nonumber \\ 
                                       & \leq \epsilon \bra \psi_j, S_1^2 S_2^2 S_3^2 \psi_j \ket \; \sup_{P} \int dq_1 \, dq_2  \frac{1} {\bra q_1 \ket^2 \, \bra q_2 \ket^2 \, \bra P - q_1 - q_2 \ket^2} \label{poinc-est1} \\
                                       & + \epsilon^{-1} \bra \phi_j, S_1^2 S_2^2 S_3^2 \phi_j \ket \; \sup_{Q} \int dp_1 \, dp_3  \frac{1} {\bra p_1 \ket^2 \, \bra Q - p_1 - p_3 \ket^{2(1-\kappa)} \, \bra  p_3 \ket^2} \label{poinc-est2},  
\end{align} 
for arbitrary $\epsilon > 0$. 
Now we apply \eqref{Cal} to bound the integral appearing in \eqref{poinc-est1} and \eqref{poinc-est2} for all $\kappa \leq 1$ when $d=1$ and for all $\kappa < 1$ if $d=2$. 
Hence \eqref{poinc-decomp}, \eqref{poinc-afterabs},  \eqref{poinc-est1} and \eqref{poinc-est2} imply: 
\begin{align} 
	|\tr \; J^{(1)} & \left( h_a(x_1 - x_2) h_a(x_1 - x_3) - \delta(x_1 - x_2) 
	\delta(x_1 - x_3)  \right) \gamma^{(3)} |\nonumber \\
                   & \leq C a^{\kappa} \left( \epsilon \tr \; J^{(1)} S_1^2 S_2^2 S_3^2 \ J^{(1)} \gamma^{(3)} + {\epsilon}^{-1} \tr \; S_1^2 S_2^2 S_3^2 \; \gamma^{(3)} \right) \nonumber \\
                   & \leq C a^{\kappa} \left( \epsilon \tr \; S_1^{-1} J^{(1)} S_1^2 J^{(1)} S_1^{-1} S_1 S_2 S_3 \gamma^{(3)} S_3 S_2 S_1 + {\epsilon}^{-1} \tr \; S_1^2 S_2^2 S_3^2 \; \gamma^{(3)} \right) \nonumber \\
                   & \leq C a^{\kappa} \left( \epsilon \| S_1^{-1} J^{(1)} S_1 \| \,  \| S_1^{-1} J^{(1)} S_1 \|  + \epsilon^{-1} \right) \tr \; S_1^2 S_2^2 S_3^2 \; \gamma^{(3)} \nonumber\\
                   & \leq C a^{\kappa} \, |\!|\!| J^{(1)} |\!|\!| \,  \tr \; S_1^2 S_2^2 S_3^2 \; \gamma^{(3)}, \label{poinc-eps}  
\end{align} 
where to obtain \eqref{poinc-eps} we choose $\epsilon =  |\!|\!| J^{(k)} |\!|\!|^{-1}$. 
Hence, the theorem is proved. 
\endprf

\subsection*{Acknowledgments}
We thank S. Klainerman and N. Tzirakis for very useful discussions.
We would like to 
thank the referee for many useful comments.  
In particular, we are grateful to Aynur Bulut and the referee for pointing out to us 
the proof of Theorem \ref{thm-Duhamel-iterate-bd-0-dim1}, 
which simplifies an earlier version.  
T.C. is much indebted to H.T. Yau and L. Erd\"os for discussions
about closely related topics around 2002 at the Courant Institute. 
The work of T.C. is supported by NSF grant DMS 0704031 / DMS-0940145.
The work of N.P. is supported by NSF grant number DMS 0758247 
and an Alfred P. Sloan Research Fellowship.

\end{document}